\documentclass{article}
\usepackage{arxiv}

\usepackage{cite}
\usepackage{amsmath,amssymb,amsfonts}
\usepackage[linesnumbered,ruled,vlined]{algorithm2e}
\usepackage{graphicx}
\graphicspath{{Images\_iVAMS-3-0/}}
\usepackage{xcolor}
\usepackage{caption}
\usepackage{subfigure}
\usepackage{wrapfig}
\usepackage{textcomp}
\usepackage{xcolor}
\usepackage{lipsum} 
\usepackage{cite}
\usepackage{multirow}
\usepackage{url}
\usepackage{amsmath}
\usepackage[compact]{titlesec}

\usepackage{braket}
\usepackage{rotfloat}
\usepackage[T1]{fontenc}    
\usepackage{hyperref}       
\usepackage{url}            

\usepackage{amsmath}
\usepackage{amssymb}
\usepackage{gensymb}
\usepackage{textcomp}
\usepackage{amsfonts}
\usepackage{booktabs}
\usepackage{siunitx}
\usepackage{hhline}

\usepackage{float}
\usepackage{tabularx}
\usepackage{multirow}
\usepackage{balance}

\newcommand{\orcid}[1]{\href{https://orcid.org/#1}{\textcolor[HTML]{A6CE39}{\aiOrcid}}}
\usepackage{array}
\newcolumntype{P}[1]{>{\centering\arraybackslash}p{#1}}
\newcolumntype{M}[1]{>{\centering\arraybackslash}m{#1}}

\def\BibTeX{{\rm B\kern-.05em{\sc i\kern-.025em b}\kern-.08em
		T\kern-.1667em\lower.7ex\hbox{E}\kern-.125emX}}

\def\matlab{MATLAB\textsuperscript{\textregistered}} 
\def\cadence{CADENCE\textsuperscript{\textregistered}} 

\title{\textbf{iVAMS 3.0}: Hierarchical-Machine-Learning-Metamodel-Integrated Intelligent Verilog-AMS for Ultra-Fast, Accurate Mixed-Signal Design Optimization}

\author{ 
	\href{https://orcid.org/0000-0003-2959-6541}{\includegraphics[scale=0.06]{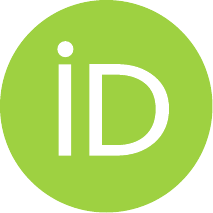}\hspace{1mm}Saraju P. Mohanty} \\
	Department of Computer Science and Engineering\\
	University of North Texas\\
	\texttt{saraju.mohanty@unt.edu} \\
	\And
	\href{https://orcid.org/0000-0002-1616-7628}{\includegraphics[scale=0.06]{orcid.pdf}\hspace{1mm}Elias Kougianos} \\
	Department of Electrical Engineering\\
	University of North Texas\\
	\texttt{elias.kougianos@unt.edu}\\
}

\begin{document}

	\makeatletter
	\maketitle

\begin{abstract}
Analog/Mixed-Signal (AMS) circuits and systems continually present significant challenges to designers with the increase of design complexity and aggressive technology scaling. This is due to the large number of design factors and parameters that must be taken into account as well as the process variations which are prominent in nano-CMOS circuits. Design optimization techniques that account for process variations while presenting an accurate and fast design flow which can perform design optimization in reasonable time are still lacking. Even with techniques such as metamodeling  that aid the design phase, there is still the need to improve them for accuracy and time cost. As a trade-off of the accuracy and speed, this paper presents a process-variation aware design flow for ultra-fast variability-aware optimization of nano-CMOS based physical design of analog circuits. It combines a Kriging bootstrapped Artificial Neural Network (ANN) metamodel with a Particle Swarm Optimization (PSO) based algorithm in the design optimization flow. The Kriging bootstrapped ANN metamodel provides a trade-off between analog-quality accuracy and scalability and can be effectively used for large and complex AMS circuits while capturing the correlation in process variations. Kriging captures correlated process variations of the circuits and accurately trains the ANN to generate the metamodels. The proposed technique uses Kriging to bootstrap target samples used for the ANN training. This introduces Kriging characteristics, which account for correlation effects between design parameters, to the ANN. The effectiveness of the design flow is demonstrated using a 180nm CMOS based PLL as a case study with as many as 21 design parameters. It is observed that the bootstrapped Kriging metamodeling is 24$\times$ faster than simple ANN metamodeling. The layout optimization for such a complex circuit can be performed effectively in a short time using this approach. The optimization flow could achieve significant reductions in the mean and standard deviation of the PLL characteristics. Thus, the proposed research is a major contribution to design for cost. 
\end{abstract}

\keywords{Metamodeling, Machine Learning, Geostatistics, Kriging,  Bootstrap Techniques, Artificial Neural Networks (ANN), Phase-Locked Lopp (PLL), Nano-CMOS, Process Variation, Mixed-Signal Circuit, Particle Swarm Optimization}

\section{Introduction}
\label{sec:Introduction}


The development and improvement of  metamodeling, (or surrogate) techniques have been gradually increasing in recent years. Significant research has been published on various metamodeling techniques for nano-CMOS applications \cite{Mohanty_TVLSI_2014Apr, Mohanty_IEEE-TSM-2012Feb}. The goal has been to develop accurate metamodels with lower computational time costs. Extensive research work exploring polynomial, artificial neural network (ANN) and Kriging techniques have been presented in \cite{GaritselovTSM2012, Okobiah2011}. ANNs are appealing because of their high accuracy and relative time efficiency. However, with the aggressive scaling of integrated circuit design, the number of design and process parameters that must be taken into consideration for design space exploration also increases. 

The design of Analog/Mixed signal (AMS) systems continues to present significant challenges. Especially in design optimization, considerable time must be spent on exploring the design space to achieve optimal designs that are robust and tolerant to the effects of process variation. This required design time is however infeasible in the face of current time-to-market constraints. For example the simulation of a PLL with full parasitics can take several days or weeks of SPICE runs for a complete space exploration.   Metamodeling design techniques are used to aid the design process by reducing the design time while maintaining accuracy. There still is a need to improve the metamodels currently used to increase time efficiency and accuracy.

\begin{figure}[htbp]
	\begin{center}
		\centering
		\subfigure[Traditional (slow) approach]{\label{fig:non_metamodeling_based_slow}\includegraphics[width=0.60\textwidth]{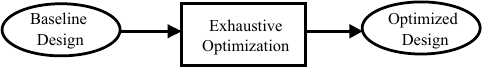}}
		\subfigure[Proposed metamodeling-based (fast) approach]{\label{fig:metamodeling_based_fast}\includegraphics[width=0.75\textwidth]{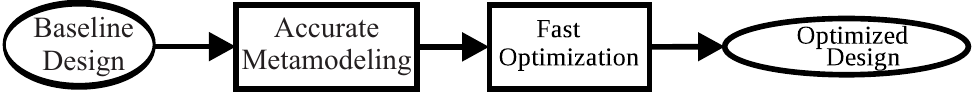}}
	\end{center}
	\caption{Fast design space exploration of analog circuits through accurate metamodeling.}
\end{figure}

In order to obtain an optimal design, a designer can optimize the actual circuit model (a SPICE netlist). This optimization on the actual circuit (Fig. \ref{fig:non_metamodeling_based_slow}) is very slow and may be even impossible for complex and nanoscale circuits with large numbers of transistors and interconnects. For fast, yet accurate design optimization of analog circuits this paper proposes the approach demonstrated in Fig. \ref{fig:metamodeling_based_fast}. In this approach, metamodels of the circuit model are first generated. The circuit optimization is then performed on the metamodels instead of the actual circuit. This makes the design exploration fast and accurate. It may be noted that \emph{metamodeling is not macromodeling}.  Macromodels are reduced complexity models but they rely on the same type of modeling and simulator as the original models (e.g., SPICE). In the \emph{metamodeling approach, the underlying system is completely decoupled from the simulator and the resulting metamodel (i.e., mathematical model of the circuit) is more general, flexible and easier to simulate and optimize than macromodels}. Macromodels are typically simplified versions of the circuit which are used in the same simulation tool and are hard to create. A metamodel has the following attributes:
\begin{enumerate}
	\item
	It is a mathematical representation of the circuit output.
	
	\item
	It is a prediction equation.
	
	\item
	Metamodels can be used in a variety of tools, such as \matlab, and are language independent.
\end{enumerate}

Kriging based techniques for generating metamodels \cite{VanBeers2005, YuICCAD2007, OkobiahVLSID2012, Hailong2009} take into account the correlation between process and circuit parameters and also incorporate a stochastic component that mitigates the deterministic nature of computer simulations, hence producing a more accurate statistical representation of the modeled circuit. The disadvantage of Kriging is that each point is predicted with a set of unique weights leading to time inefficient metamodel generations on large design spaces.  ANN generated metamodels on the other hand are more time efficient for simulations. In this paper we propose a metamodeling based design approach that combines the benefits of Kriging with the accuracy and time efficiency of ANN models to produce accurate metamodels which are also more effectively process aware. Kriging is used to bootstrap the design samples used for training the ANN models, thus introducing a process aware component into the training set. We show that the Kriging trained ANN models are more process aware accurate than the bare ANN models. 

We also present a process aware design flow that incorporates into different levels of the design process techniques to account for the effects of process variation. It combines a Kriging bootstrapped ANN metamodeling technique with a Particle Swarm Optimization (PSO) algorithm \cite{Jin2007, Coello2004} in the design optimization flow. The effectiveness of the design flow is shown using a PLL as a case study. PSO techniques are part of genetic, evolutionary, and population based algorithms. In using PSO to train ANN models, the PSO aims to optimize the set of design parameters that are fed into the training of the ANN models. This modification hence improves the selection of parameters from the training set, thus resulting in a faster and more efficient training of the ANN models. While the case-study has been presented with PSO optimization in the current paper, the use of other similar optimization algorithms is possible. While PSO and Neural Network (NN) metamodels have demonstrated increased accuracy \cite{Vilovic2009}, certain design factors such as device parameter variations continue to pose a significant concern to circuit performance estimation. Analog circuits are particularly sensitive and hence prone to these effects \cite{KuoICS2008}.

Finally, this work should be placed within the general framework of Digital Twins (DT) in the context of the semiconductor space \cite{Synopsys_Digital_Twin}. DTs are models that encompass the entire life cycle of an electronic product, such as, but not limited to, integrated circuits (IC). Our work is specific to IC design and as such could form a component of a DT but would not supersede it.

This paper is organized as follows: Section \ref{sec:Novel_Contributions} highlights the novel contribution of this work.  Section \ref{sec:Related_Research} contains  a brief discussion on current related research. The proposed design optimization flow methodology is  presented in section \ref{sec:PSO_optimization}. The Kriging-bootstrapped metamodeling process is described in Section \ref{sec:bootstrapKriging}. Section  \ref{sec:psoAlgorithm} discusses the optimization algorithm used.  Experimental results are presented in section \ref{sec:results}. Finally, in section \ref{sec:Conclusion} a summary conclusion and future research ideas are presented.

\section{Novel Contributions of the Current Paper}
\label{sec:Novel_Contributions}

The overall contribution of this paper is an ultra-fast, accurate statistical design exploration flow that combines Kriging bootstrapped neural network metamodels and particle swarm optimization to advance the state-of-art in design for cost. This is due to significant reduction in the design cycle time which leads to decrease in non-recurrent design cost of the chip. This paper presents the following \emph{novel contributions} to the state-of-the art of analog/mixed-signal CAD:
\begin{enumerate}
\item
Fast and accurate physical design and optimization flow incorporating process awareness in analysis, characterization and optimization of performance measures.
\item
Process-variations aware accurate and scalable metamodeling using Kriging bootstrapped Neural Networks.
\item
Adaptation of the PSO algorithm for nano-CMOS based process-variation aware optimization.
\item
A case study exploration using a 180 nm CMOS based PLL design.
\end{enumerate}
It may be noted that a generic overview of Kriging metamodeling is presented in \cite{Okobiah_ISVLSI2014}. A Kriging metamodel approach for process variation analysis is presented in \cite{Okobiah_ISQED_2014}. The current paper presents a natural progression of our research to ultra-fast physical design optimization of large analog blocks through the use of Kriging metamodeling.

\section{Related Prior Research}
\label{sec:Related_Research}

Related research in the context of this work includes the design and formulation of modeling and metamodeling techniques to improve the accuracy of such models and simultaneously reduce design exploration time. Polynomial regression methods which include response surface methodology (RSM) \cite{Zakerifar2009, Biles2007, Dellino2009} are one of the most common and reliable methods explored. However, low order polynomial regression techniques are not very accurate for global design space exploration \cite{Ankenman2008, Aggarwal2007}. They assume a random error between design variables while in the presence of process variations, these errors may be correlated, especially in deep nanometer process technology. Non-polynomial based metamodels, particularly built from Neural Network  training have also been reported to surpass polynomial regression techniques \cite{LWang2005, Khosravi2009, Zobel2008, Sabuncuoglu2002}. NN techniques use a learning process to continuously train weights used in approximating these models. The weight training process is critical in the development of NN models and  research in exploring techniques for optimizing this process is currently active. A technique popularly used is applying optimization algorithms to optimize the weight training of NN models \cite{Vilovic2009}. Use of Kriging training for the NN architecture provides a trade-off between the accuracy of Kriging and scalability of the NN method \cite{Okobiah_ISQED_2014}.  In the current paper, we propose to infuse the characteristics of Kriging based techniques by bootstrapping the sample data points which are then used for the NN training process. We believe that the bootstrapped  data points enhance the modeling of process variation effects.

Monte Carlo (MC) simulations methods have been a reliable and effective method for yield analysis of designs. Multiple simulations of the modeled circuit are run while varying the design and process parameters (transistor length, transistor width, supply voltage, thickness oxide, etc.) to reflect the effect of process variations. In \cite{KuoICS2008} hierarchical statistical analysis and regression based techniques have been explored for variation analysis. The proposed Kriging bootstrapped NN model is analyzed for statistical variation using the MC method.

\section{Process Variation Aware Ultra-Fast Design Optimization Flow for Mixed-Signal Circuits}
\label{sec:PSO_optimization}

\begin{figure*}[htbp]
	\centering
	\includegraphics[width=0.98\textwidth]{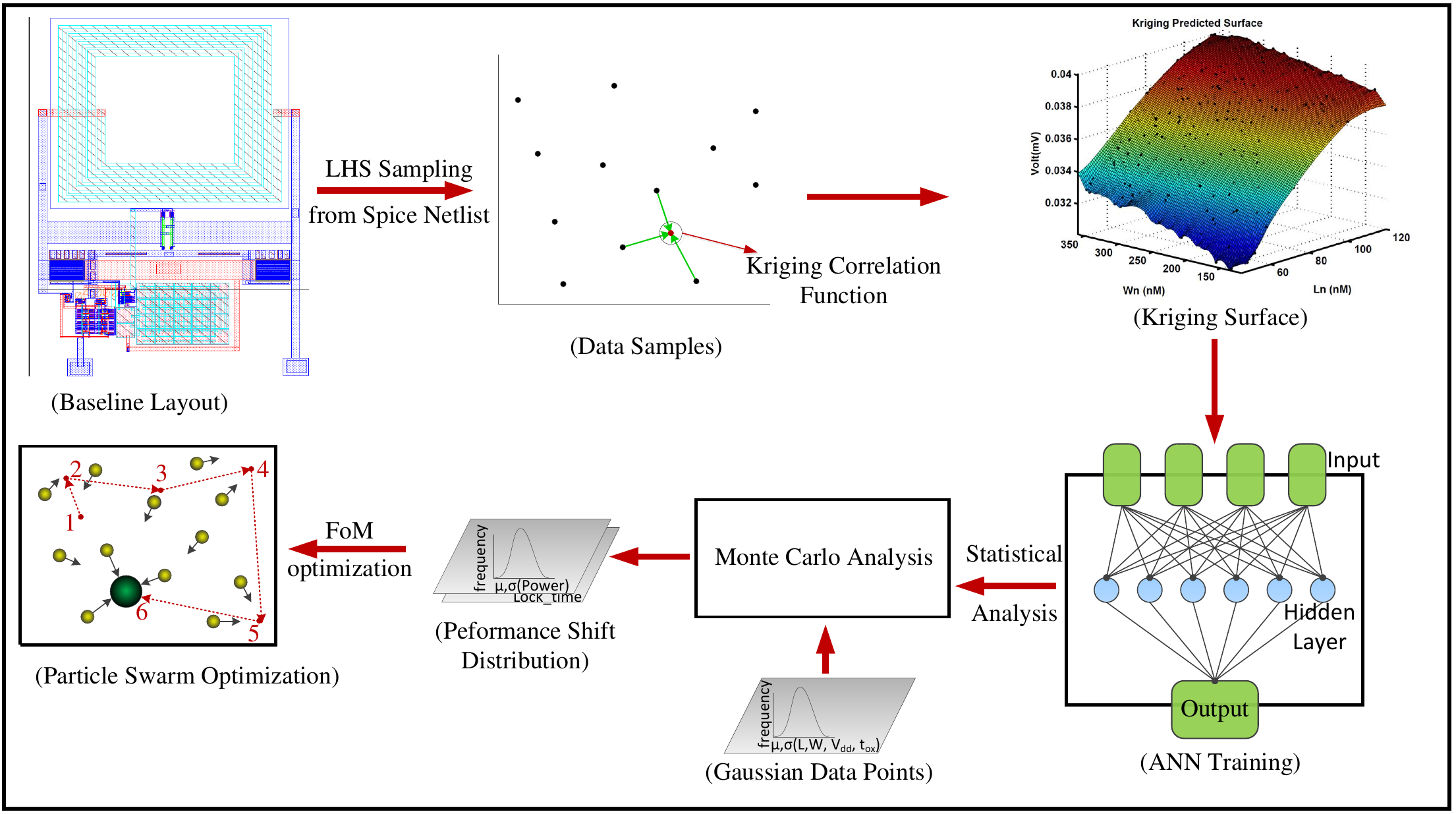}
	\caption{Proposed high level design flow with Hierarchical Machine Learning Modeling \cite{Mohanty_ISQED_2015}. }
	\label{Fig:High-Level-Flow}
\end{figure*}

We propose a novel design flow that integrates a Kriging bootstrapped metamodeling process with the PSO algorithm for the design optimization of nano-CMOS circuits as, depicted in Fig. \ref{Fig:High-Level-Flow}. The key idea is to generate a Kriging surface using a small number of analog simulations with latin hypercube sampling (LHS) of the variables. An NN architecture is then trained to create a metamodel of the baseline circuits. Statistical analysis and optimization is performed over the metamodel instead of its SPICE netlist. The use of metamodels for design optimization iterations significantly speeds up the design-optimization process and analog-level accuracy is maintained by the use of accurate metamodels which are generated from the parasitic-aware netlist.

\begin{figure}[hptb]
\centering
\includegraphics[width=0.65\textwidth]{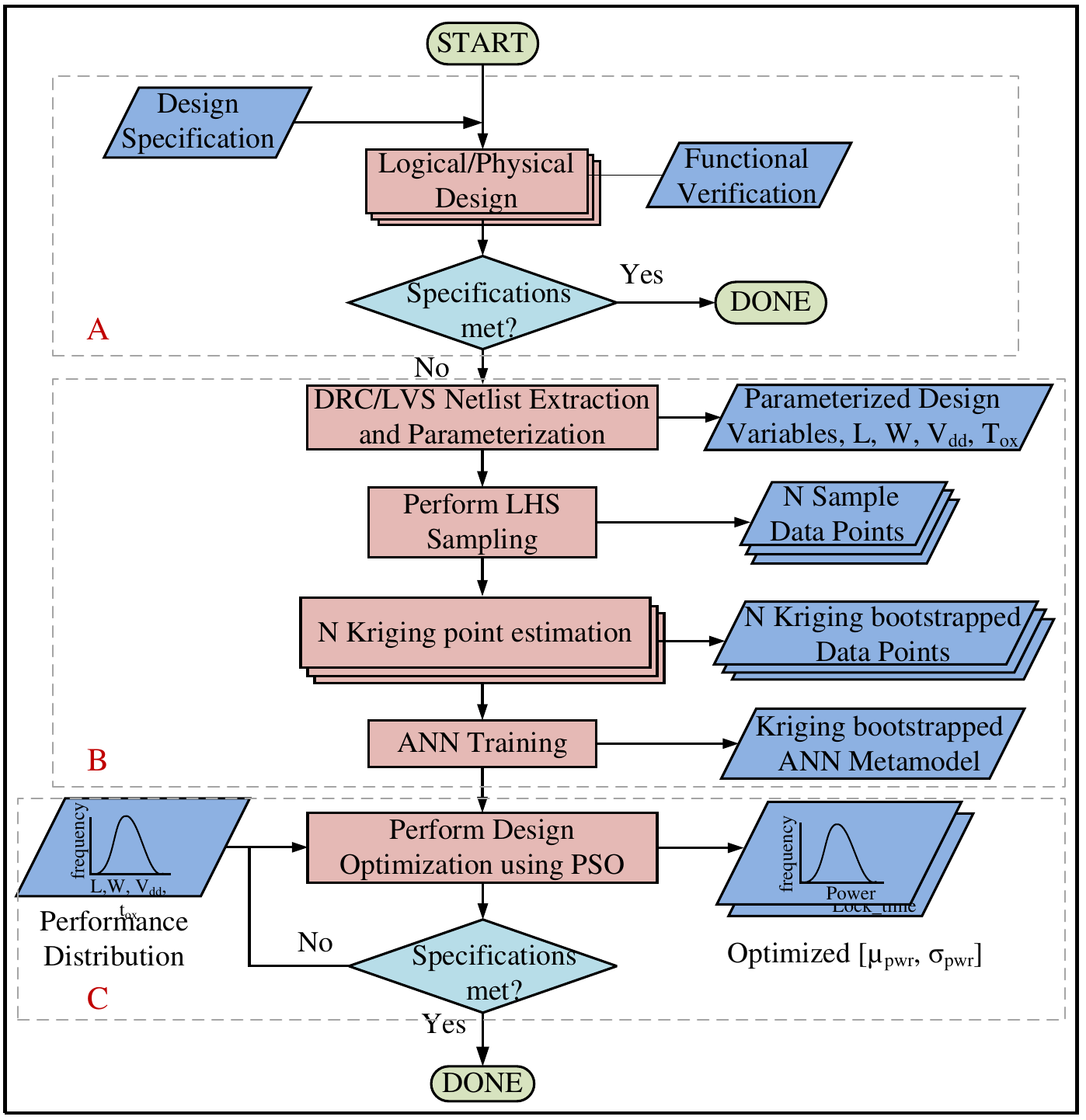}
\caption{Proposed design optimization flow \cite{Mohanty_ISQED_2015}.}
\label{Fig:overall_flow}
\end{figure}

The overall flow of the design process shown in Fig. \ref{Fig:overall_flow} highlights the major phases of the design flow. The first phase labeled ``A'' consists of the baseline logical and physical design. In this phase, the baseline design is drawn both as a circuit schematic and the associated layout. The baseline is simulated for functional verification of the performance objectives. The functional verification also serves to characterize the circuit design objectives which are defined in Section \ref{Sec:Case_Study}.  The next phase involves the creation of the process variation aware metamodel of the circuit design. The first step in this phase is the identification and parameterization of the variables used to create the metamodel from the extracted parasitic netlist. Incorporating the process parameters early on in the design phase ensures a process variation aware metamodel. An LHS of the circuit from the parasitic netlist is then used by Kriging techniques to bootstrap the sample data points infusing process variation characteristics. We detail this process in Section \ref{sec:bootstrapKriging}. The Kriging bootstrapped points are used for the NN Training. The final phase is the process aware design optimization. The optimization algorithm  is used together with the created metamodel and design objectives as an input to optimize the design. The final design parameters are then used to update the physical design for an optimal design of the circuit. The process aware design optimization phase is discussed in detail in Section \ref{sec:psoAlgorithm}.

The use of bootstrapping Kriging with ANN improves both the accuracy and speed of the design flow process. Bootstrapping Kriging techniques ensure process aware accuracy while ANN metamodeling techniques have been shown to be very fast. The overall design flow metamodel incorporates process variation awareness in the design metamodeling phase and the design optimization phase.

\section{Proposed Kriging Bootstrapped Artificial Neural Network (ANN) Metamodeling}
\label{sec:Kriging_ANN_metamodeling}

\begin{figure}[htbp]
	\center
	\includegraphics[width=0.75\textwidth]{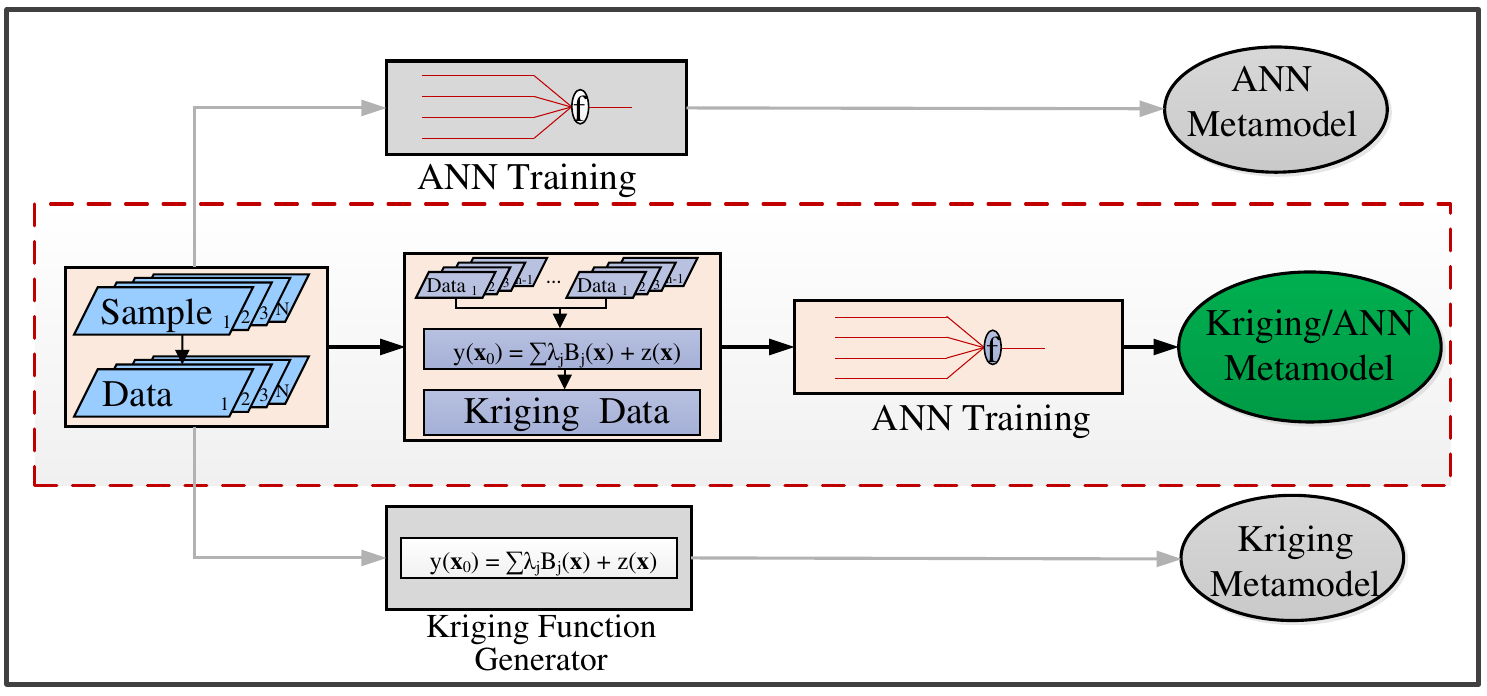}
	\caption{Proposed Kriging Bootstrapped ANN Metamodel Generation Flow \cite{Okobiah_ISQED_2014}.}
	\label{Fig:metamodel_Flow}
\end{figure}

In this section, we introduce and discuss the proposed Kriging bootstrapped Artificial Neural Network metamodeling technique. First we briefly introduce traditional Kriging and Artificial Neural Network metamodeling and then discuss our proposed modifications. Our proposed kriging bootstrapped ANN metamodeling technique is shown in Fig. \ref{Fig:metamodel_Flow}.

\subsection{Kriging Based Metamodeling}
\label{subsec:krigingMetamodel}

Kriging prediction techniques were originally applied in geostatistics but have since been explored for other applications such as circuit design \cite{OkobiahVLSID2012, Hailong2009,YuICCAD2007}. Kriging metamodeling combines  polynomial regression with a stochastic approach to mitigate the deterministic nature of computer simulations. The Kriging equations can be expressed in the form of the following:
\begin{equation}
	y(\mathbf{x_0})=\sum_{j=1}^L \lambda_j B_j(\mathbf{x}) + z(\mathbf{x}),
\end{equation}
where $y(\mathbf{x_0})$, is a stochastic function which predicts the response $y$ at the design point $(\mathbf{x_0})$. $\{B_j(\mathbf{x}),j=1,\cdots,L\}$ is a specific set of basis functions over the design domain $D_N$ and $\lambda_j$ are fitting coefficients (also known as weights) to be determined based on the Kriging method applied. $z(\mathbf{x})$ is
a stochastic process with zero mean and is based on a spatial correlation function.

In calculating the weights $\lambda_j$ for estimating Kriging functions, the autocorrelation between the input parameters is accounted for and characterized by the covariance function of the following form \cite{Bohling2005}:
\begin{equation}
	r(\mathbf{s},\mathbf{t})=\mathrm{Corr}(z(\mathbf{s}),z(\mathbf{t})).
\end{equation}

This property of Kriging prediction is exploited to model the effects of process variation on circuit metamodels. The correlations between the process variation of the design and process parameters are taken into consideration in calculating the weights for the metamodel functions. The drawback of Kriging is that the weight for each predicted point is unique and involves matrix calculations which could become time intensive for a large design space.

\subsection{Artificial Neural Network Metamodeling}
\label{subsec:neuralMetamodel}

ANN models consist of simple computational elements with rich interconnections between the elements. They are modeled after biological neural networks which operate in a parallel and distributed fashion. The neural networks create models over a set of inputs by training the weights of the interconnections. Multilayer and radial neural networks are few of the commonly employed neural networks. The multilayer network which is used in this work uses a combination of non-linear activation functions in a hidden layer and a linear activation function in the output layer. The linear layer of the function output can be expressed as follows:
\begin{equation}
	{v}_{i}=\sum _{i=1}^{s}{w}_{ji}{x}_{i}+{w}_{{j}^{0}},
\end{equation}
where ${w}_{ji}$ is the weight of the connection between the $j$th element in the hidden layer and the $i$th component in the input layer $x_i$ and ${w}_{{j}^{0}}$ is a constant bias \cite{Garitselov2012b}. The input layer is represented using a sigmoid function such as the following:
\begin{equation}
	{b}_{j}\left({\nu }_{i}\right)=\tanh\left(\lambda {v}_{j}\right) .
\end{equation}

The neural network utilizes an algorithm (a training function) that updates the weights and biases of the interconnections to minimize the error between the predicted point and the actual response. For this work, the ANN metamodel was created using a \matlab toolbox which implements the Levenberg-Marquardt optimization algorithm \cite{MATLAB}.

\subsection{Kriging Bootstrapped ANN Metamodeling}
\label{subsec:krigingneuralMetamodel}

Metamodeling techniques based on Kriging prediction  have been explored in \cite{OkobiahVLSID2012, Biles2007}. In estimating performance points, Kriging prediction techniques take into account the correlation effects between design parameters.  This characteristic is very appealing and can be used to model the correlation effects between design parameters due to process variation for design processes deep in the nanometer range. The drawback to Kriging based techniques is that the weights used for each point prediction are unique and have to be calculated for each performance point to be estimated using linear algebra calculations (mostly matrix inversion). This can lead to potential time consuming metamodel generation for high dimensional designs and very large design spaces. Artificial Neural Network (ANN) training, which has also been presented for NanoCMOS metamodeling in \cite{Garitselov2012b}, has been shown to be robust and accurate for high dimensional models \cite{LWang2005}. While ANN also produces highly accurate models, it does not effectively model process variation effects with correlations present.

Hence, the proposed metamodeling technique aims to combine Kriging and ANN to generate accurate models which account for the effects of correlated process variation in a fast and efficient manner. Fig. \ref{Fig:metamodel_Flow} highlights the already presented methods for ANN and Kriging metamodel generation. For each method sample data points are generated using a Latin Hypercube Sampling (LHS) design and then are either fed into an ANN trainer or a Kriging function generator. In the proposed metamodel generation method, the sample data points are fed into a Kriging generator that produces an intermediate set of sample data points (bootstrapped) which are then fed into the ANN trainer. This method feeds the ANN trainer Kriging generated sample data points which are process and correlation aware. We demonstrate that using the Kriging generated sample data points will result in a more robust metamodel which is process variation aware and also less time intensive.

\begin{figure}[hptb]
	\center
	\includegraphics[width=0.55\textwidth]{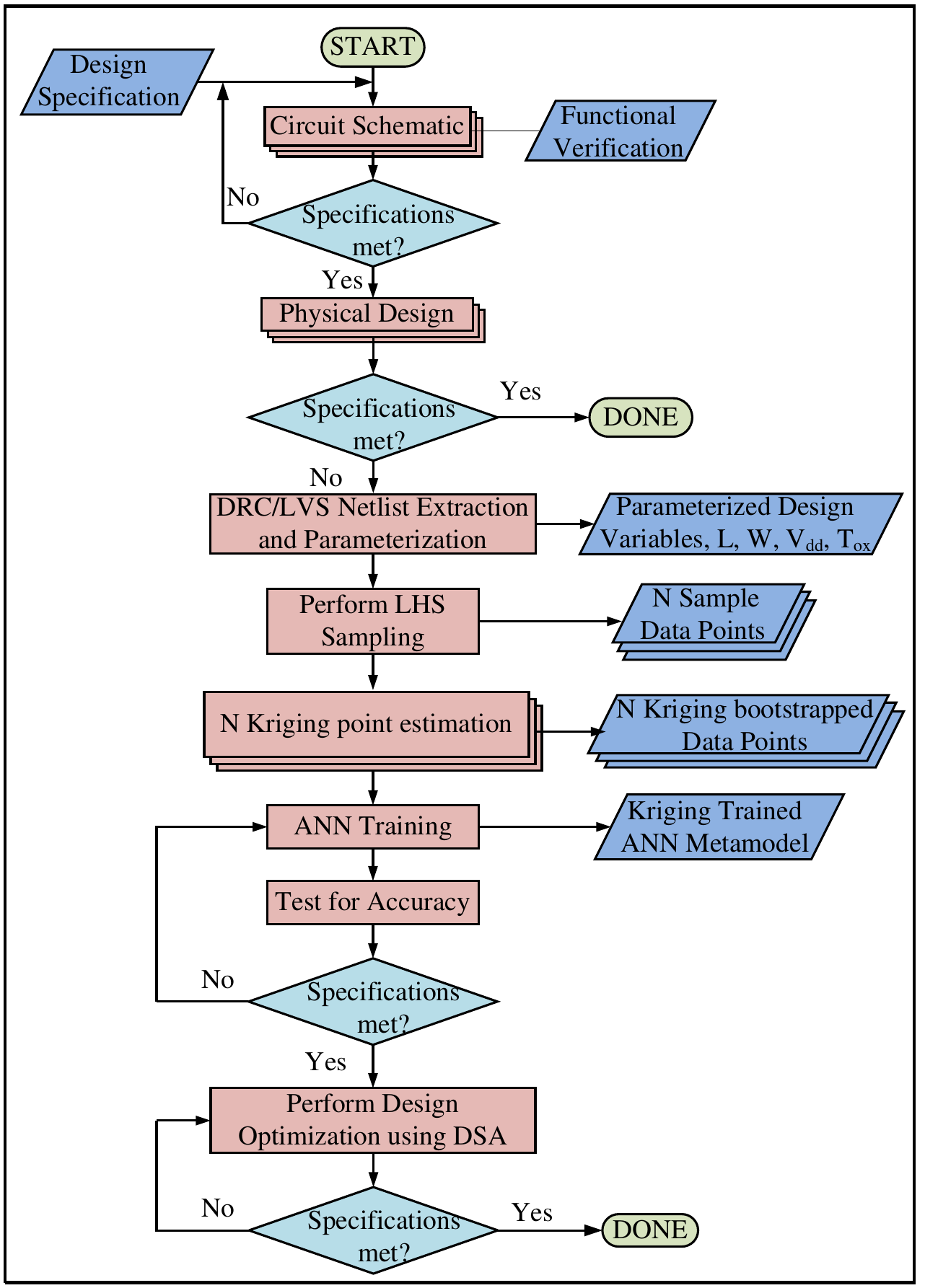}
	\caption{Proposed Metamodel Design Flow \cite{Okobiah_ISQED_2014, Mohanty_ISQED_2015}.}
	\label{Fig:design_flow}
\end{figure}

The methodology for the generation of the proposed metamodel-based design flow is shown in Fig. \ref{Fig:design_flow}. The first step involves creating a SPICE netlist of the design. The functional simulation of the circuit schematic is performed to ensure the SPICE model meets design specifications. The physical layout design is also constructed using Design Rule Check (DRC) and Layout vs. Schematic (LVS) verification to ensure a match to the circuit schematic. The physical layout design is used to generate a silicon-aware accurate model (netlist). The performance of the physical design is often degraded due to the parasitic effects. A fully extracted parasitic netlist, including resistance ($R$), capacitance ($C$) and self ($L$) and mutual inductance ($K$) is used to ensure silicon-level accuracy.

The generation of the metamodel is based on the extracted parasitic $RCLK$ netlist. In order to generate data sample points, the extracted parasitic netlist is parameterized for the design and process variables and then simulated to eliminate the strenuous task of physically varying the design parameters on the physical layout design. The Latin Hypercube Sampling technique is used in the proposed method to vary the design and process parameters. LHS methods generate $N$ random sample points from a given design space. They divide the design space into equal intervals and then randomly select design points from an interval in such a way that each interval appears once in a row-column matrix of the design space. Several techniques may be used to select the data points including uniformly, midpoints or randomly. We use Random LHS which has been reported to generate more accurate models \cite{GaritselovTSM2012}. The LHS parameter points are used as inputs to the parameterized netlist to generate corresponding performance outputs (data point) for each sample point.

The next step in the metamodeling process is the Kriging bootstrapping of the data points. The generated sample points are  fed into a Kriging metamodel generator. We implement the Kriging metamodel presented in our previous work [reference removed] for this process.  We generate $N$ Kriging bootstrapped data points by using $N-1$ points and the Kriging method to estimate the $N$th point. $N$ iterations of this process will generate $N$ Kriging bootstrapped data points which are then used for the ANN training.

The ANN training process is used to create metamodels for each performance objective (Figure-of-Merit or FoM) characterized for the design. In this research, 4 metamodels were created for the Phase Locked Loop (PLL) circuit described in Section \ref{Sec:Case_Study}.

The final step of the metamodel design flow is the verification and test of accuracy of the generated metamodel. The statistical metric used to verify the accuracy is the Root of Mean Square Error (RMSE). The expression of the RMSE is given as follows:
\begin{equation}
	RMSE  =  \sqrt{\frac{1}{N}\sum_{i=1}^N\left( Y_i - \widehat{Y}_i\right)^2},
\end{equation}
where $N$ is the number of sampled points, $Y_i$ is the ``true'' circuit response (SPICE simulation results) and $\widehat{Y}_i$ is the metamodel predicted response. The RMSE measures the difference between the metamodel and the SPICE model where a smaller value indicates a more accurate model.

\section{The Proposed Process-Variation Aware Kriging Bootstrapped Metamodeling}
\label{sec:bootstrapKriging}

The metamodeling technique incorporates Kriging to infuse process variation characteristics to the sampled data.  Kriging by itself has been successfully used for metamodel generation with high accuracy \cite{Biles2007}. The property of Kriging which makes it very appealing and lends to its high accuracy is its ability to take into account the correlation between the input parameters in performance point prediction. This can be effectively utilized to model the correlation between the process parameters which also serves as input in the sample data point bootstrapping.

In Kriging, the weights are chosen to minimize the variance under the unbiasedness constraint that $E(\widehat{y}(x) - y(x)) = 0$, where  $\widehat{y}(x)$ is the predicted response at point $x$ and $y(x)$ is the true response. Hence the weights are chosen so that the following expression is satisfied:
\begin{equation}
	\label{eqn:Krigingweights}
	\sum_{j=1}^n \lambda_j = 1.
\end{equation}
The weights then are given by the following:
\begin{equation}
	\label{eqn:Krigingweights_equation}
	\left(\!\!\begin{array}{c}
		\lambda_1\\
		\vdots\\
		\lambda_n\\
		\mu\end{array}  \!\!\right)
	= \Gamma ^{-1}\left(\!\!\begin{array}{c}
		\gamma(e_1,e_0)\\
		\vdots\\
		\gamma(e_n,e_0)\\
		1
	\end{array}  \!\!\right),
\end{equation}
where $\mu$ is a Lagrange multiplier used to ensure equation (\ref{eqn:Krigingweights}). $\Gamma$ is the covariance matrix of the observed points and for ordinary Kriging is given by:
\begin{equation}
	\Gamma = \left( \!\! \begin{array}{cccc}
		\gamma(e_1,e_1)  & \cdots & \gamma(e_1,e_n) & 1 \\
		\vdots           & \ddots & \vdots          & 1 \\
		\gamma(e_n,e_1)  & \cdots & \gamma(e_n,e_n) & 1 \\
		1               &   1    &   1             & 0
	\end{array} \!\!\right),
\end{equation}
where
\begin{equation}
	\gamma(e_1, e_2) = E \left( |z(e_1) - z(e_2)|^2 \right).
\end{equation}

A disadvantage of Kriging is that it uses a set of matrix equations in calculating the unique weights for point predictions. For large circuits and high dimensional designs, the time cost can become expensive. The use of NN on the other hand can generate metamodels which are ultra-fast and robust in accuracy. The NN models however do not efficiently model the effects of process variation. To ensure accuracy and time efficiency as well, we present a Kriging bootstrapped metamodeling technique that combines the accuracy of Kriging with the speed of NN models.

The metamodel generation process takes in sample data from the extracted parasitic netlist. The sample data points are fed into the Kriging metamodel generator for resampling of the data (bootstrapping). We generate $N$ Kriging bootstrapped data points by using $N-1$ points and the Kriging method to estimate the $Nth$ point. $N$ iterations of this process will generate $N$ Kriging bootstrapped data points which are then used for the NN training. The NN training process is used to create metamodels for each performance objective (Figure-of-Merit or FoM) for the design. In this work, 4 metamodels were created for the PLL circuit described in section \ref{sec:results}. 

\section{Particle Swarm Optimization (PSO) Algorithm for Process-Variation Aware Optimization}
\label{sec:psoAlgorithm}

This Section presents a detailed discussion on the proposed particle swarm optimization (PSO) algorithm which performed statistical design exploration over the bootstrapped Kriging metamodels. 

\subsection{Particle Swarm Optimization (PSO) Algorithm}

PSO is a type of evolutionary swarm intelligence algorithm for numerical optimization problems. Swarm intelligence algorithms are based on the exploitation of social or communal behavior of naturally or artificially occurring agents to collectively search for solutions.  While heuristic in nature and based on social behaviors, swarm intelligence algorithms have proved to be very effective in optimization \cite{PoliJAEA2008, Civicioglu2012, Dorigo2006}, and circuit design \cite{BennourDTIS2010, GaritselovGLSVLSI2012}. 

PSO uses candidates of solutions termed ``particles'' and modeled after movement of organisms in  bird flocks or fish schools, hence the term swarm.  The PSO algorithm models the swarm like motion to implement a collective search algorithm, where the particles correspond to search agents which explore different habitats based on the quality of previous solutions. The quality of the results is expressed through the position and velocity of the particles. The particle's movement are updated based on its previous solution (local intelligence) and are also influenced by the global best known solution. Reiterative updates of the particles swarm towards the best solution.


\subsection{Process-Variation Aware Adaptation of Particle Swarm Optimization (PSO) Algorithm}

\begin{figure}[hbpt]
\center
\includegraphics[width=0.65\textwidth]{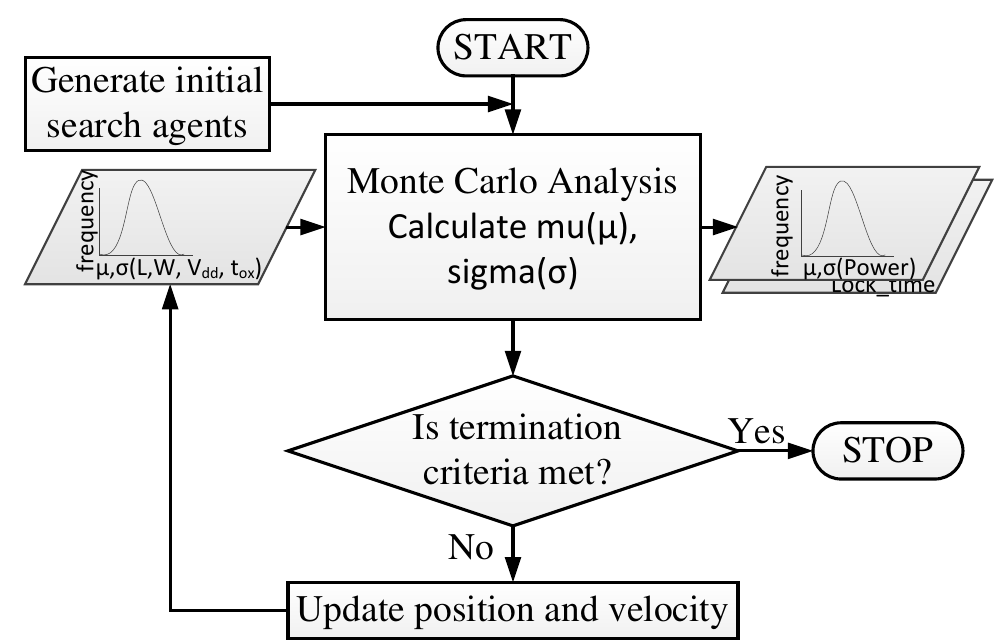}
\caption{Flow diagram for the PSO algorithm \cite{Mohanty_ISQED_2015}.}
\label{Fig:algo_Flow}
\end{figure}

The optimization problem implemented in this flow is to minimize the power consumption of the PLL circuit using the locking time as a design constraint. The process aware optimization of the circuit involves minimizing the mean $\mu$ and standard deviation $\sigma$ of the optimal power consumption. As a specific example, the optimization function can be expressed as follows:
\begin{equation}
\text{Minimize}[{\mu }_{pwr} + 3 {\sigma }_{pwr}],
\end{equation}
while subjected to locking time constraint. The PSO algorithm for the PLL is shown in Algorithm \ref{Alg:PSO_Algorithm}. Fig. \ref{Fig:algo_Flow} presents an illustration of the algorithm.

\begin{algorithm}[htbp]
\caption{Particle Swarm Optimization of PLL over the Metamodels \cite{Mohanty_ISQED_2015}.}
\label{Alg:PSO_Algorithm}
\KwIn{Tuning parameter set $X$, Bootstrapped Kriging metamodels, Tuning parameter ranges.}
\KwOut{$X=(x_1, x_2,...,x_n)$ parameter set with optimized statistical performance;}
\Begin{
\SetKwHangingKw{HData}{SET:}
\HData{$N$, number of particles}
\HData{Max$_{iteration}$, counter $\longleftarrow$ 0}
\HData{local best $l_{x_i} \longleftarrow$ current position}
\HData{global best $g_{x_i} \longleftarrow$ current position}
\SetKwHangingKw{HData}{Initialize:}
\HData{weight for swarm effect $\varrho$}
\HData{velocity for swarm effect $w$}
\While{counter $<$ Max$_{iteration}$}{
   \ForEach{N}{
   Monte Carlo analysis over metamodels with nominal $x_i$\;
   $v_i$ = $wv_i$ + $\varrho_p\tau_p(lx_i-x_i)$ + $\varrho_g\tau_g(gx_i-x_i)$\;
   $x_i$ $\leftarrow$  $x_i$ + $v_i$\;
   \If{$x_i$ $<$ $l_{x_i}$}{
  	  $l_{x_i}$ $\leftarrow$ $x_i$\;	
      \If{$l_{x_i}$ $<$ $g_{x_i}$}{
        $g_{x_i}$ $\leftarrow$ $l_{x_i}$\;
        }
     }
  }
}
}
\KwResult{Parameter set $X$ with minimized $\mu$, $\sigma$;}
\end{algorithm}

%


\section{Case Study Circuit: A 180nm CMOS PLL}
\label{Sec:Case_Study}

\begin{figure}[hptb]
	\centering
	\includegraphics[width=0.65\textwidth]{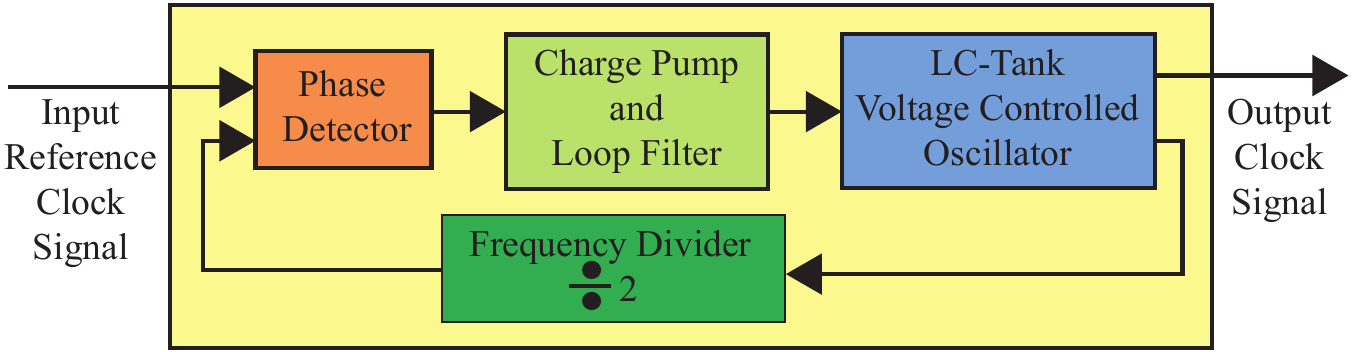}
	\caption{High level system diagram for the PLL.}
	\label{Fig:PLL_systemLevel}
\end{figure}

The phase locked loop (PLL) is a closed feedback loop circuit system whose output signal is locked to a reference input signal. The PLL is a critical component in many Analog/Mixed Signal (AMS) systems including processors, telecommunication devices, Field-Programmable Gate Arrays (FPGAs), controllers and many other systems. The system level diagram of a PLL  shown in Fig. \ref{Fig:PLL_systemLevel} shows the major components of the PLL which include the phase detector, the charge pump/loop filter, the voltage controlled oscillator (VCO) and the frequency divider.

The reference clock feeds the input signal to the phase detector, which compares and detects the phase difference between the input signal and the output from the VCO. The charge pump generates a supply charge in proportion to the error detected in the phase difference. The generated signal is then filtered by the loop filter to produce a control signal which the VCO uses to produce an output signal which is locked to the reference input signal. The divider is an optional component of the PLL which is used to generate an output signal which is a multiple of the reference input signal.

\begin{figure}[hptb]
	\centering
	\includegraphics[width=0.30\textwidth]{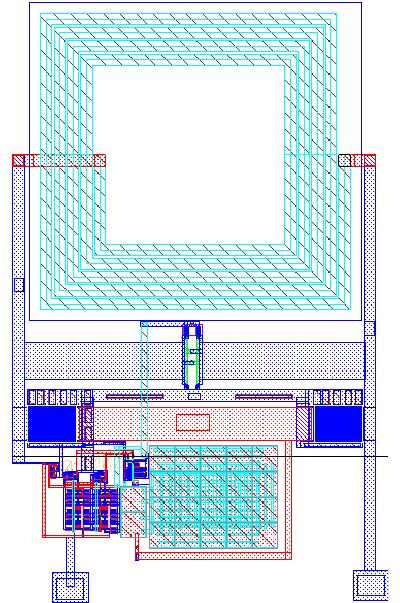}
	\caption{Physical design of the 180 nm PLL \cite{Okobiah_ISQED_2014}.}
	\label{Fig:PLL_physicalDesign}
\end{figure}

The schematic and physical layout design of the PLL using a 180 nm CMOS technology was produced on the \cadence Virtuoso platform. Figure \ref{Fig:PLL_physicalDesign} shows the physical layout of the design.

The PLL was characterized for  power consumption, frequency output, locking time and jitter. The baseline design values are shown in Table \ref{Table:baselinePLL}. The FoMs selected are Power ($P_{PLL}$), Frequency ($F_{PLL}$), Locking time ($Lck_{PLL}$), and Jitter ($J_{PLL}$). The design objective is the minimization of power consumption using the locking time as optimization cost and 21 parameters as design variables.

\begin{table}[htpb]
	\centering
	\caption{Characterization of PLL for Figures of Merit (FoM) \cite{Okobiah_ISQED_2014}.}
	\label{Table:baselinePLL}
	\begin{tabular}{|l|c|c|c|c|}
		\hline
		PLL   & Power & Frequency & Locking & Jitter \\
		Circuit   & $P_{PLL}$ & $F_{PLL}$ & $Lck_{PLL}$ & $J_{PLL}$ \\
		\hline \hline
		Layout       & 2.48 mW     &  2.66 GHz   & 5.51 $\mu$s   & 16.80 ns      \\
		\hline
	\end{tabular}
\end{table}

\section{Process Variation Aware Statistical Analysis}
\label{sec:Process_variation}

In this section we perform a process variation aware statistical analysis of the generated Kriging trained ANN metamodel. Monte Carlo simulation experiments are a common method for the analysis of process variation on analog circuits in order to estimate the yield and efficiency of the design. Monte Carlo analysis enables an efficient investigation of the design space by randomly generating a distribution test case of design variables. The set of test cases form a given probability distribution with a mean of the nominal value of the variable. This is particularly efficient in high dimensional designs where a test case simulation time increases exponentially. For example, in our PLL case study circuit which has 21 design and process parameters, even a high and low test case will require 2$^{21}$ simulations.

\begin{figure}[hptb]
	\center
	\includegraphics[width=0.75\textwidth]{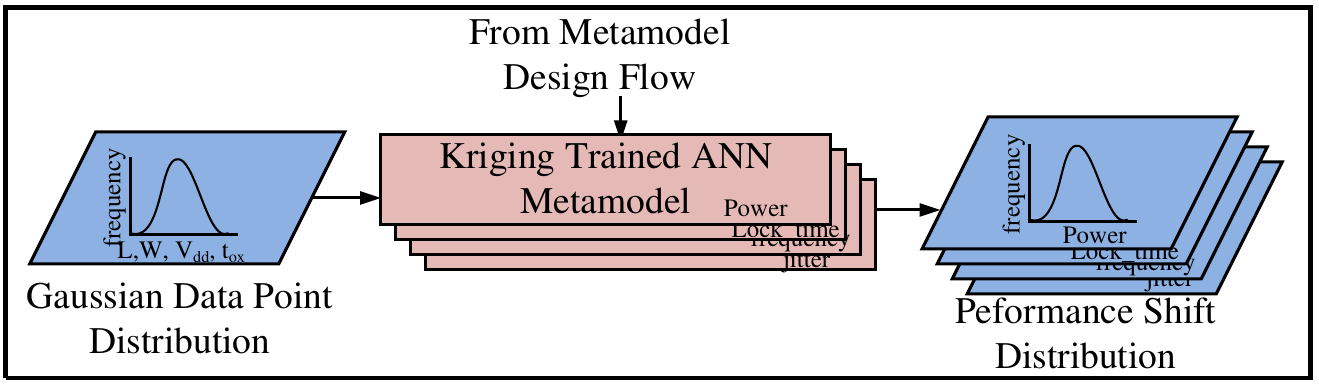}
	\caption{Statistical Variation Analysis}
	\label{Fig:statisticalVariation_flow}
\end{figure}

The selection of design and process parameters significantly affects the accuracy of the analysis. A sensitivity test is usually performed to select parameters which are most sensitive to performance measure. Reported research \cite{KuoICS2008,  KangISSC2005, Nassif2001} shows that the length ($L_n$, $L_p$), width ($W_n$, $W_p$) and oxide thickness ($T_{ox}$) have a significant effect on the performance shift. $L_n$, $L_p$, $W_n$, $W_p$ for the various sub-circuit components of the PLL have been used as design parameters. The nominal values are selected from the baseline design in Section \ref{Sec:Case_Study} and a Gaussian distribution with 10 \% standard deviation is used to generate the sample set for the metamodel simulation.  Fig. \ref{Fig:statisticalVariation_flow} summarizes the statistical analysis process. $N=1000$ Monte Carlo simulations are performed for each FoM.

The performance results of the Monte Carlo analysis are compared with an analysis from the spice simulation of the PLL design in the next section.

\section{Experimental Results}
\label{sec:results}

In this section, simulation experiments are performed on the case study 180 nm PLL design discussed in section \ref{Sec:Case_Study} to illustrate the effectiveness of our proposed approach. The Kriging bootstrapped neural network metamodel is built using the \matlab Neural Network toolbox and the mGstat toolbox \cite{MATLAB, mGstat}. The model used in the design flow is discussed in Section \ref{subsec:krigingneuralMetamodel}. The extracted parasitic netlist is parameterized and used for the sample data point generation. mGstat is used to implement the Kriging boostrapping of the sample data points and then the metamodel is generated using the Neural Network toolbox. Four metamodels are generated, one for each Figure of Merit (FoM) (Power($P_{PLL}$) , Frequency ($F_{PLL}$), Locking time ($Lck_{PLL}$), and Jitter ($J_{PLL}$)) characterizing the PLL. A Monte Carlo method is used to evaluate the statistical distribution of the four FoMs. A Gaussian distribution of 1000 samples is used for the simulation analysis. The results are presented in Fig. \ref{fig:K_ANN}. Also presented in Figures \ref{fig:Kriging}, and \ref{fig:ANN} are statistical distributions using the ANN and the Kriging based metamodels, respectively, for comparison to the proposed metamodel.

\subsection{Design Objective and Simulation Setup}
\label{sec:simulationSetup}

\begin{figure}[htbp]
\centering
\includegraphics[width=0.40\textwidth]{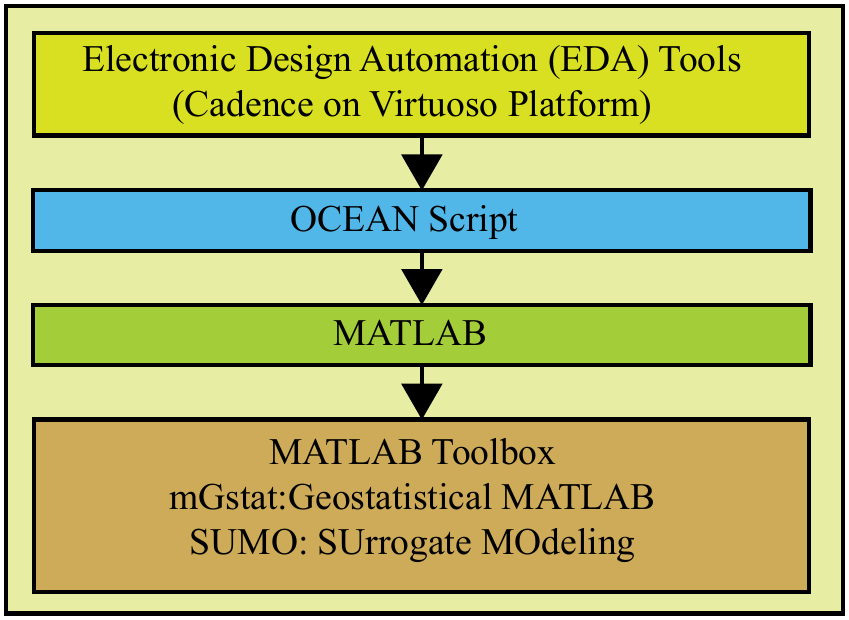}
\caption{Experimental Steps and EDA/Non-EDA Tool Interactions.}
\label{FIG:Tool_Interactions_for_Experiments}
\end{figure}

All of the logical schematic and physical layout designs were performed using the \cadence virtuoso platform. The full blown parasitic (RLCK) netlist is extracted and parameterized with respect to the corresponding design variables. The parameterized netlist is used as the circuit description for design sampling. An ocean script is created with the parameterized netlist that can automate the design sampling procedure using \matlab. The Spectre analog simulator was used to perform the simulations. The algorithm used to generate the Kriging metamodels was written using \matlab with the help of the toolboxes mGstat \cite{mGstat} and SUMO \cite{Gorissen2010}. A diagram showing the different tool interactions is shown in Fig. \ref{FIG:Tool_Interactions_for_Experiments}. Any design engineer can use this as a guideline for tool usage to reproduce our results when needed to be used in their circuit design.

The 180 nm PLL circuit design was simulated for a low power consumption optimization using the locking time as a design constraint. The optimization objective was to increase the yield and tolerance to process variation by minimizing the mean and variance of the power dissipation of the PLL. The statistical optimization process uses the Particle Swarm Optimization (PSO) algorithm described in Algorithm \ref{Alg:PSO_Algorithm} and illustrated in Fig. \ref{Fig:algo_Flow} to search the design space on the generated Kriging trained ANN metamodel. Monte Carlo simulation experiments are a common method for the analysis of process variations on analog circuits in order to estimate the yield and efficiency of the design.  Monte Carlo enables an efficient analysis of the design by randomly generating a distribution test case of design variables. The set of test cases is derived from a given probability distribution with a mean of the nominal value of the variable. The initial nominal values are selected from the baseline design and a Gaussian distribution with 10 \% standard deviation is used to generate the sample set for the metamodel simulation. Subsequent iterations of the algorithm give the nominal point for the Monte Carlo simulations.

\subsection{Results Analysis}
\label{sec:resultsAnalysis}

Table \ref{Table:krigingNeuralAccuracy} shows the accuracy of the proposed Kriging Bootstrapped Trained ANN Metamodels. The Root Mean Square Errors (RMSE) for each of the FoMs is shown. A lower value of RMSE indicates a higher accuracy. The low RMSE values thus demonstrate that the created metamodels are sufficiently accurate and can be used for design exploration.

\begin{table}[hptb]
	\centering
	\caption{ Statistical Accuracy of Kriging Generated Points \cite{Okobiah_ISQED_2014}.}
	\label{Table:krigingNeuralAccuracy}
	\begin{tabular}{|l|c|}
		\hline
		FoM                     & RMSE          \\
		\hhline{|=|=|}
		Power ($P_{PLL}$)         & 2.51 x 10 $^{-6}$       \\
		\hline
		Frequency ($F_{PLL}$)     & 5.68 x 10 $^{-13}$       \\
		\hline
		Locking Time($Lck_{PLL}$) & 5.01 x 10 $^{-12}$        \\
		\hline
		Jitter ($Lck_{PLL}$)      & 1.69 x 10 $^{-19}$   \\
		\hline
	\end{tabular}
\end{table}

The Monte Carlo results for the various metamodels are shown in Table \ref{Table:ProcessResults}. A Monte Carlo analysis on the
SPICE model is used as baseline to compare the results. The results are also compared with the bare Kriging and ANN metamodels.

\begin{table*}[hptb]
	\centering
	\caption{Statistical Analysis for Accuracy of Neural Network Metamodel for PLL FoMs \cite{Okobiah_ISQED_2014}.}
	\vspace{-5pt}
	\begin{tabular}{|l|c|c|c|c|c|c|c|c|}
		\hline
		\multicolumn{2}{|l|}{}    &\multicolumn{1}{c|}{Circuit}  & \multicolumn{2}{c|}{Kriging-ANN} & \multicolumn{2}{c|}{Kriging} & \multicolumn{2}{c|}{ANN}\\ \cline{3-9}
		\multicolumn{2}{|l|}{}             & Value       & Value       & error (\%) & Value       & error (\%) & Value       & error (\%) \\
		\hhline{|==|=|=|=|=|=|=|=|} 
		\multirow{2}{*}{$P_{PLL}$ }  & Mean & 2.48 mW     & 2.40 mW     & 3.22       & 2.50 mW     & 0.81       & 2.50 mW     & 0.81      \\
		\cline{2-9}
		& STD  & 0.42 mW     & 0.34 mW     & 19.05      & 0.51 mW     & 21.43      & 0.69 mW     & 64.28     \\
		\hline
		\multirow{2}{*}{$F_{PLL}$}   & Mean & 2.66 GHz    & 2.51 GHz    & 5.64       & 2.66 GHz    & 0.11       & 2.74 GHz    & 5.38       \\
		\cline{2-9}
		& STD  & 10.95 MHz   & 41.93 MHz   & 282.92     & 3.72 MHz    & 66.03      & 51.9 MHz    & 373.97       \\
		\hline
		\multirow{2}{*}{$Lck_{PLL}$} & Mean & 5.51 $\mu$s & 5.11$\mu$s  & 7.26       & 5.51 $\mu$s & 0.07       & 5.20 $\mu$s & 5.63      \\
		\cline{2-9}
		& STD  & 0.72 $\mu$s & 0.44 $\mu$s & 38.88      & .58 ns      & 10.25      & 1.01 $\mu$s & 40.27      \\
		\hline
		\multirow{2}{*}{$J_{PLL}$}   & Mean & 16.80 ns    & 14.69ns     & 10.25      & 16.78ns     & 0.12       & 17.91 ns    & 6.61      \\
		\cline{2-9}
		& STD  & 1.32 ps     & 4.50 ps     & 240.91     & 0.68ps      & 48.48      & 19.17 ps    & 1352.22      \\
		\hline
	\end{tabular}
\label{Table:ProcessResults}
\end{table*}

\begin{figure}[hptb]
	\centering
	\subfigure[Mean] {\label{fig:mc_mean}\includegraphics[width=0.45\textwidth]{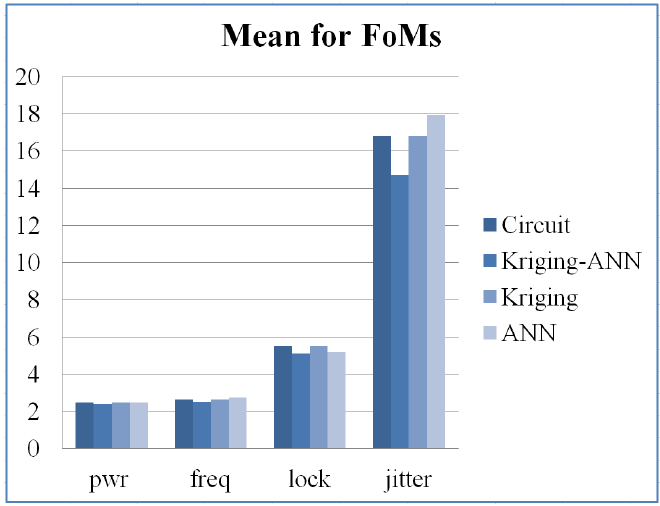}}
	\subfigure[STD] {\label{fig:mc_STD}\includegraphics[width=0.45\textwidth]{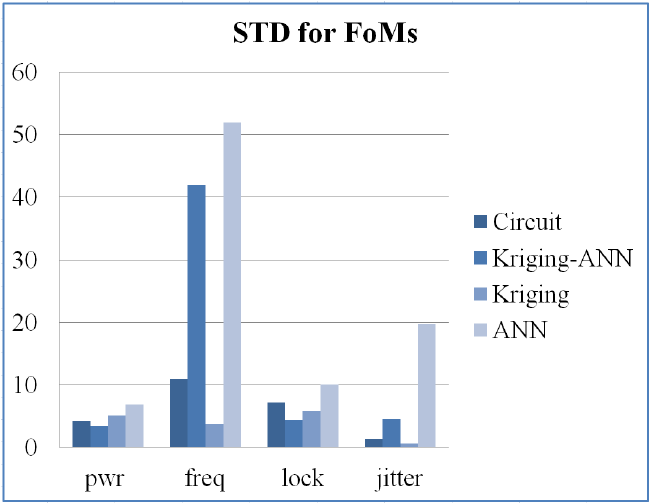}}
	\caption{Comparative Results with Kriging and Neural Network \cite{Okobiah_ISQED_2014}.}
	\label{fig:mcMean_STD}
\end{figure}

\begin{figure*}[hptb]
	\centering
	\subfigure[Power] {\label{fig:mc_pwrKANN}\includegraphics[width=0.45\textwidth]{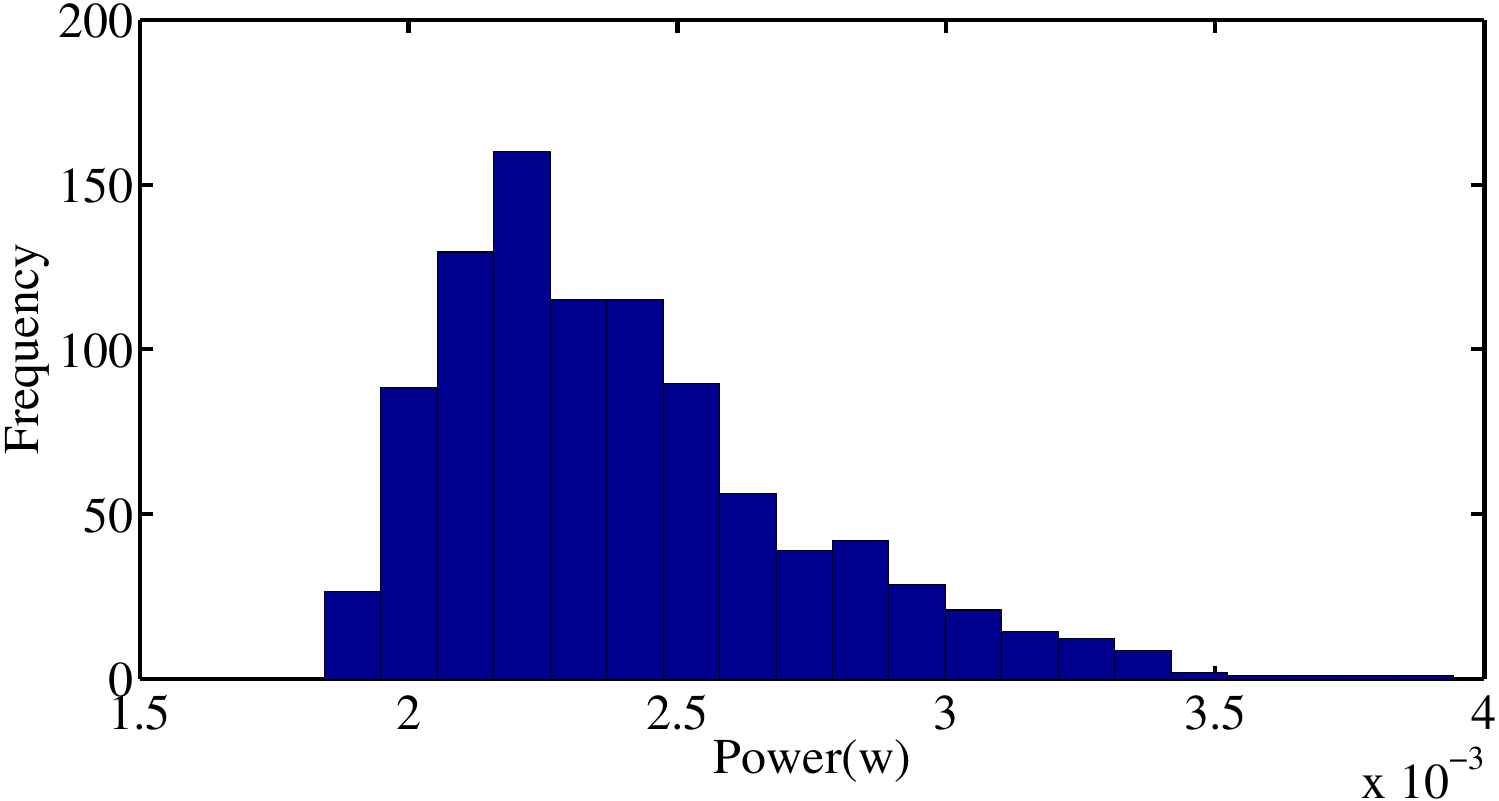}}
	\subfigure[Frequency] {\label{fig:mc_freqKANN}\includegraphics[width=0.45\textwidth]{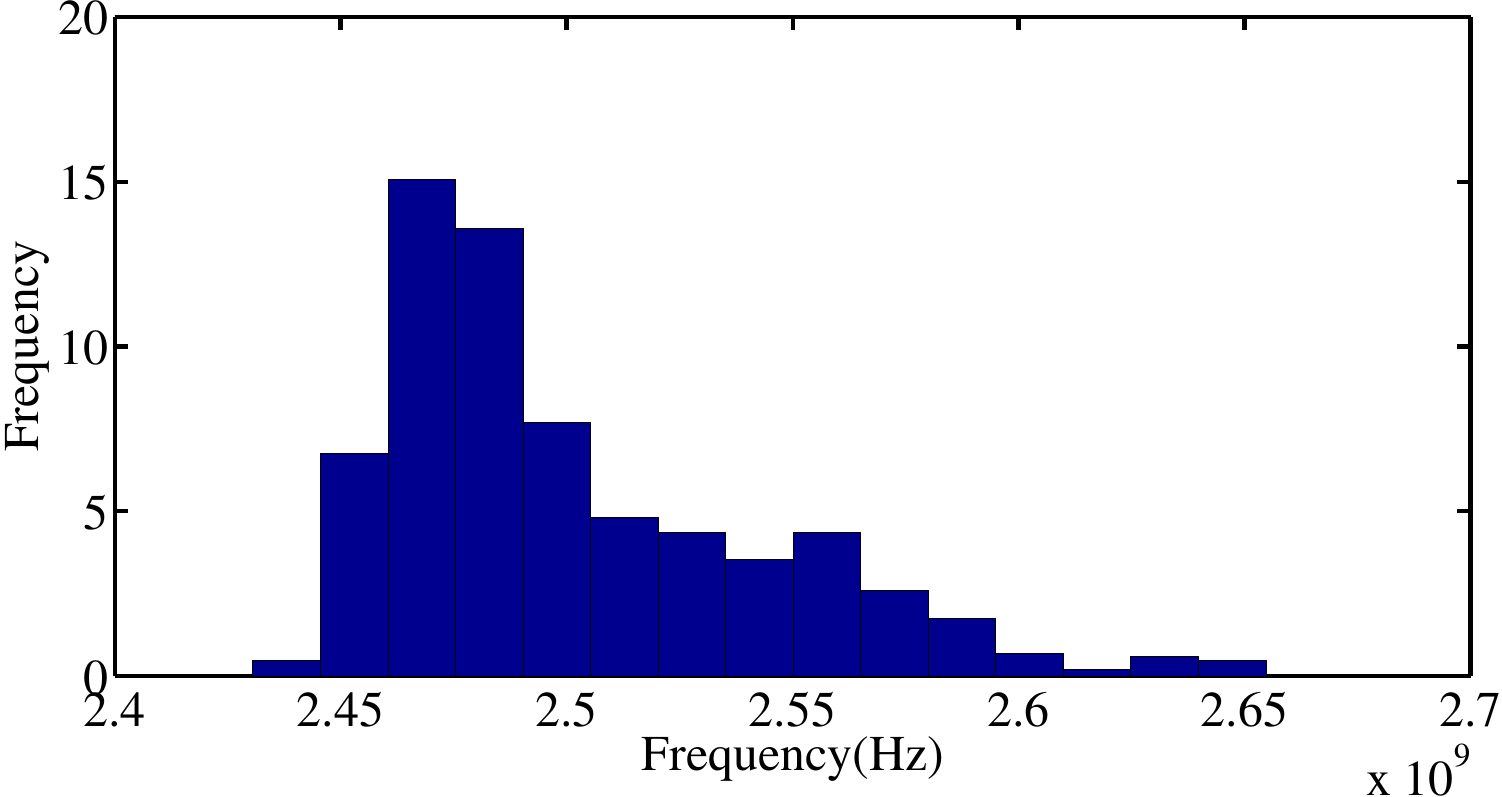}}
	\subfigure[Locking Time] {\label{fig:mc_lockKANN}\includegraphics[width=0.45\textwidth]{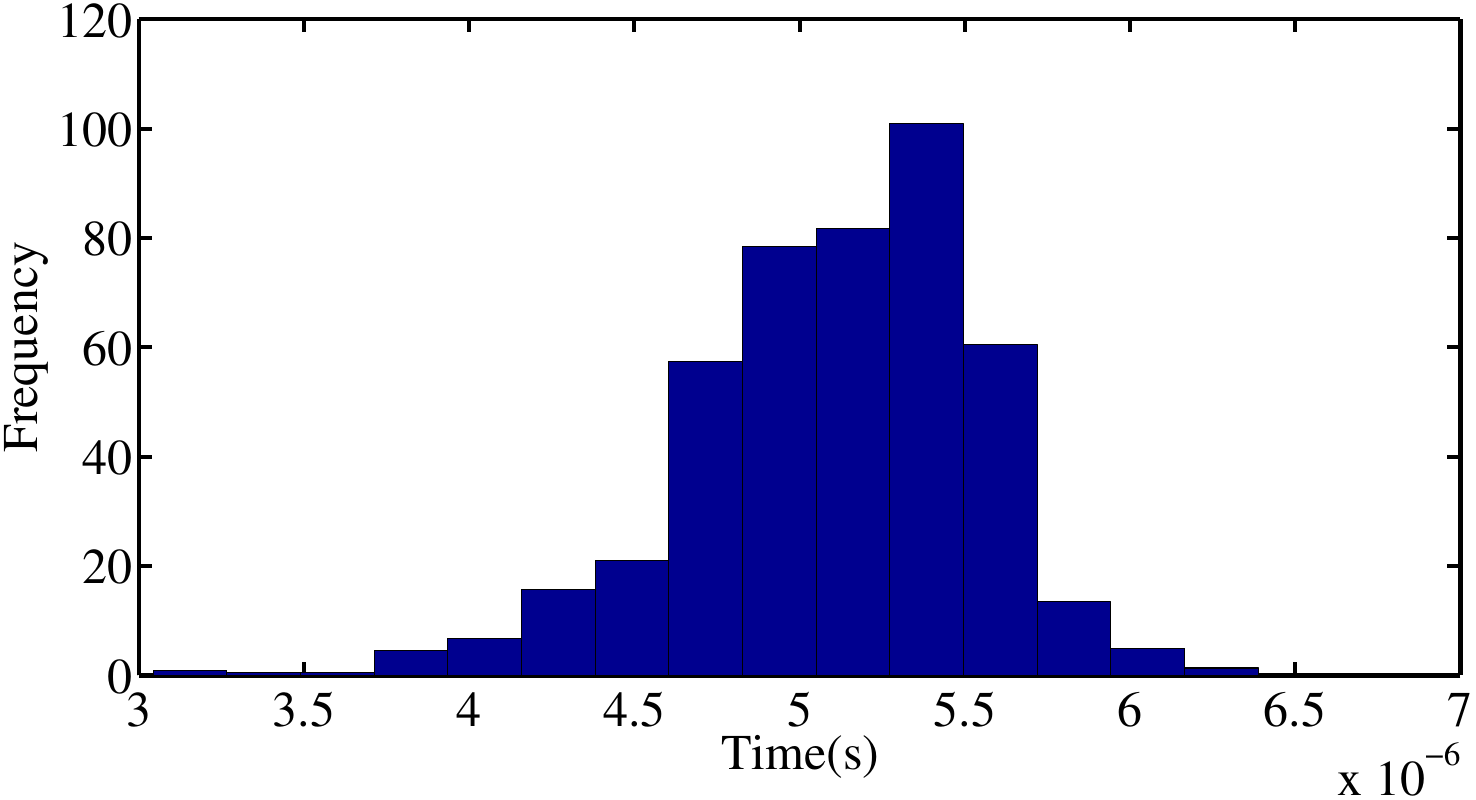}}
	\subfigure[Jitter] {\label{fig:mc_jittKANN}\includegraphics[width=0.45\textwidth]{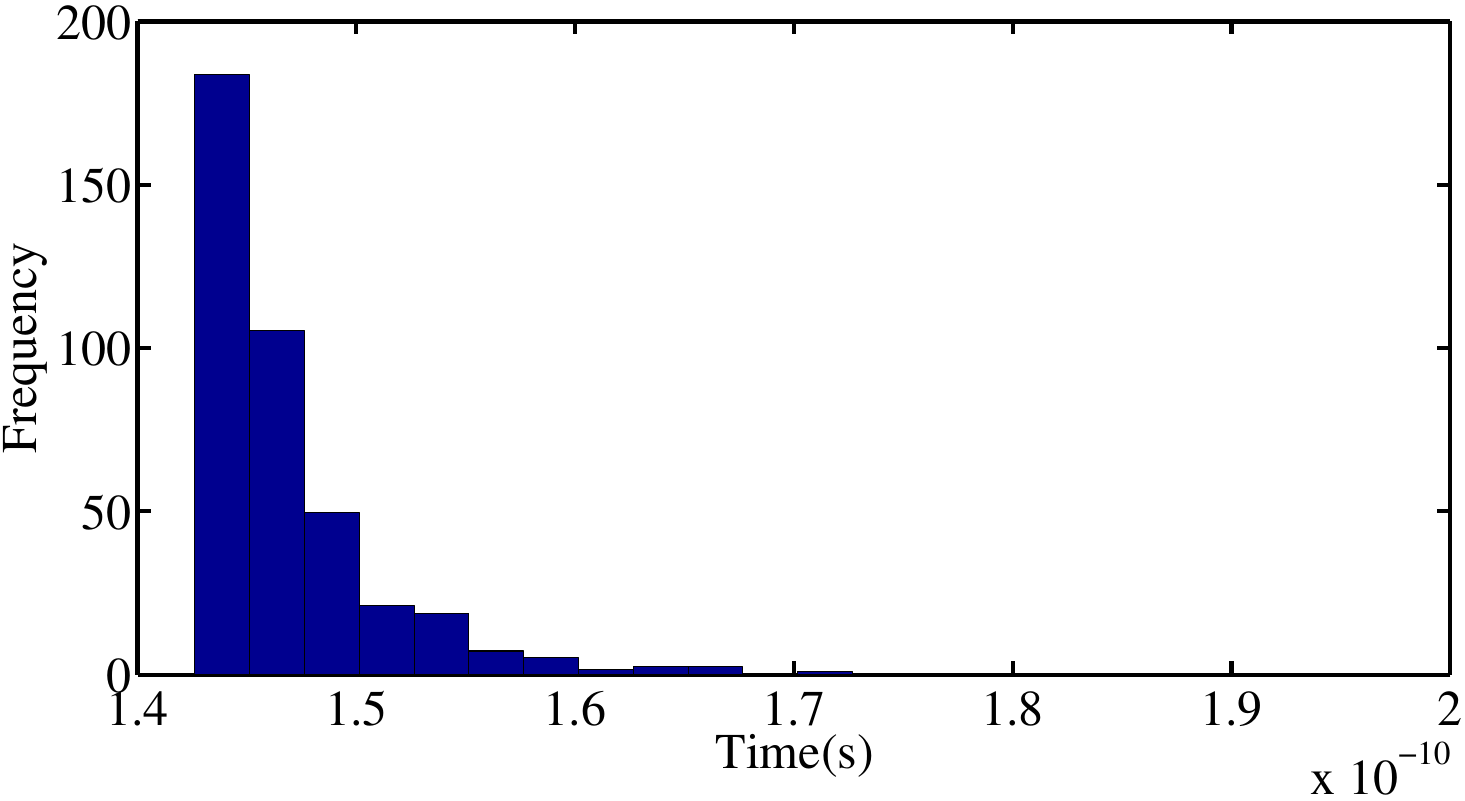} }
	\caption{Statistical Analysis of FoMs using Kriging Bootstrapped Trained Neural Network based metamodeling \cite{Okobiah_ISQED_2014}.}
	\label{fig:K_ANN}
\end{figure*}

\begin{figure*}[hptb]
\centering
\subfigure[Power] {\label{fig:mc_pwrKrig}\includegraphics[width=0.45\textwidth]{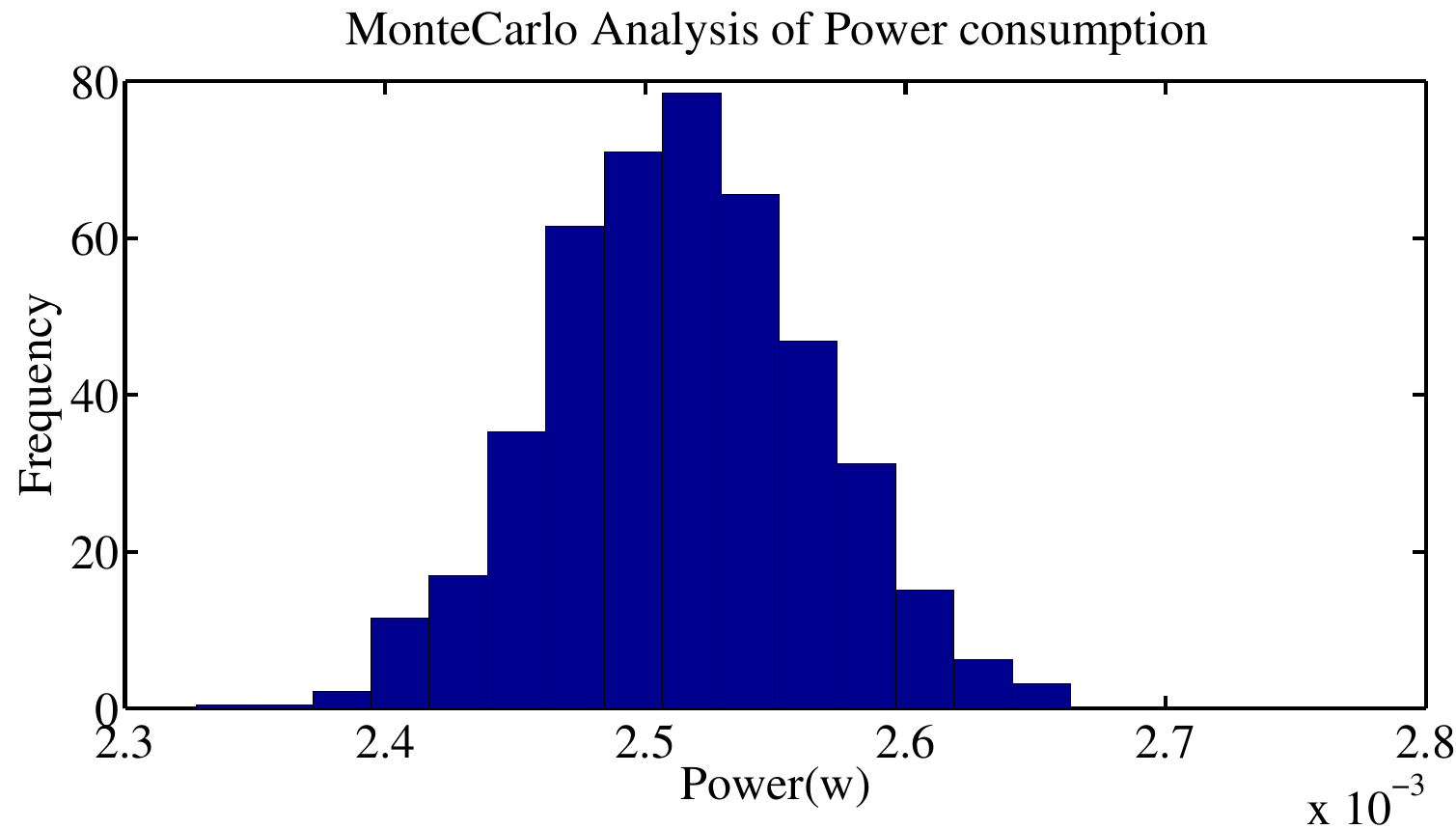}}
\subfigure[Frequency] {\label{fig:mc_freqKrig}\includegraphics[width=0.45\textwidth]{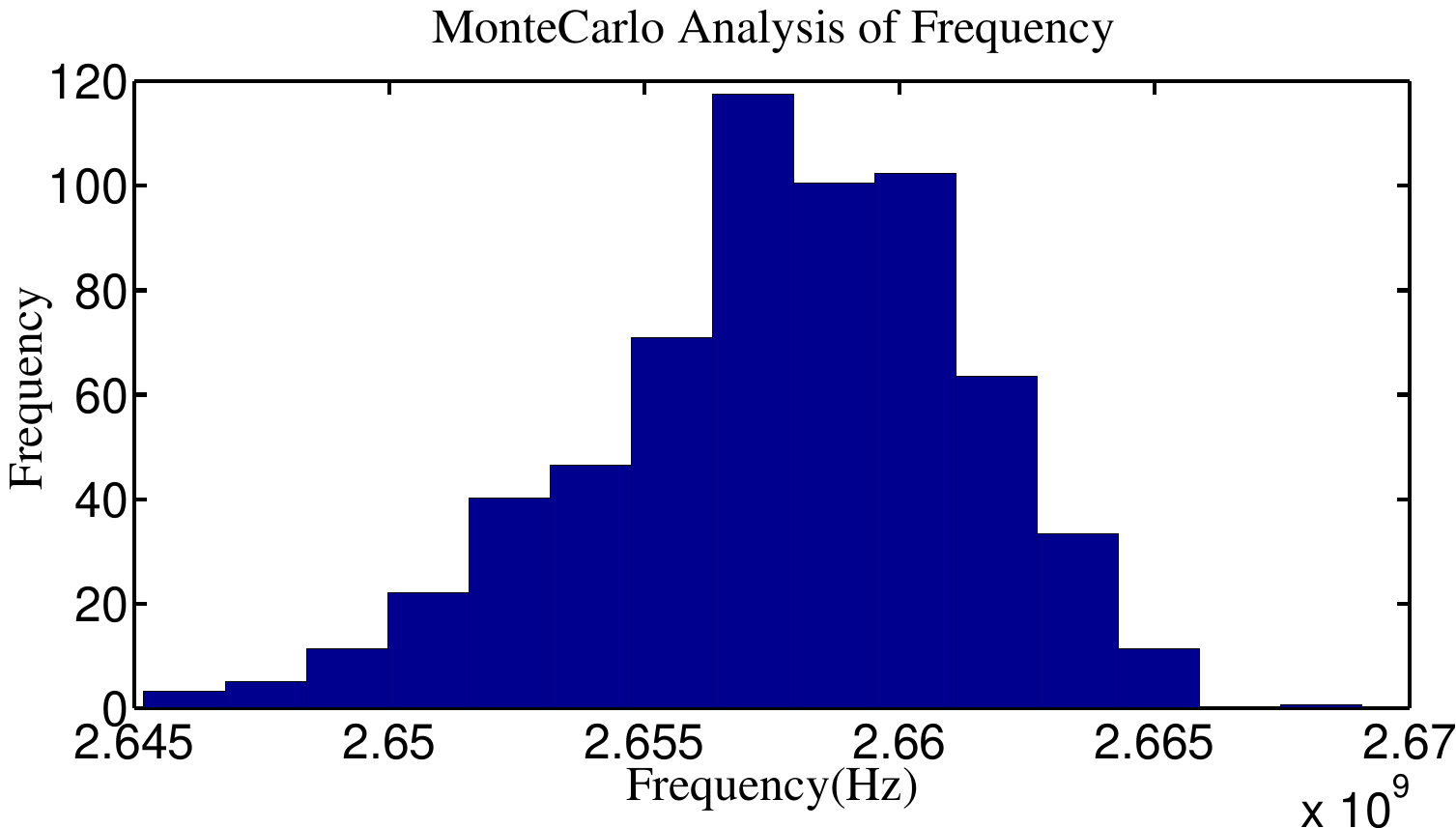}}
\subfigure[ Locking Time] {\label{fig:mc_lockKrig}\includegraphics[width=0.45\textwidth]{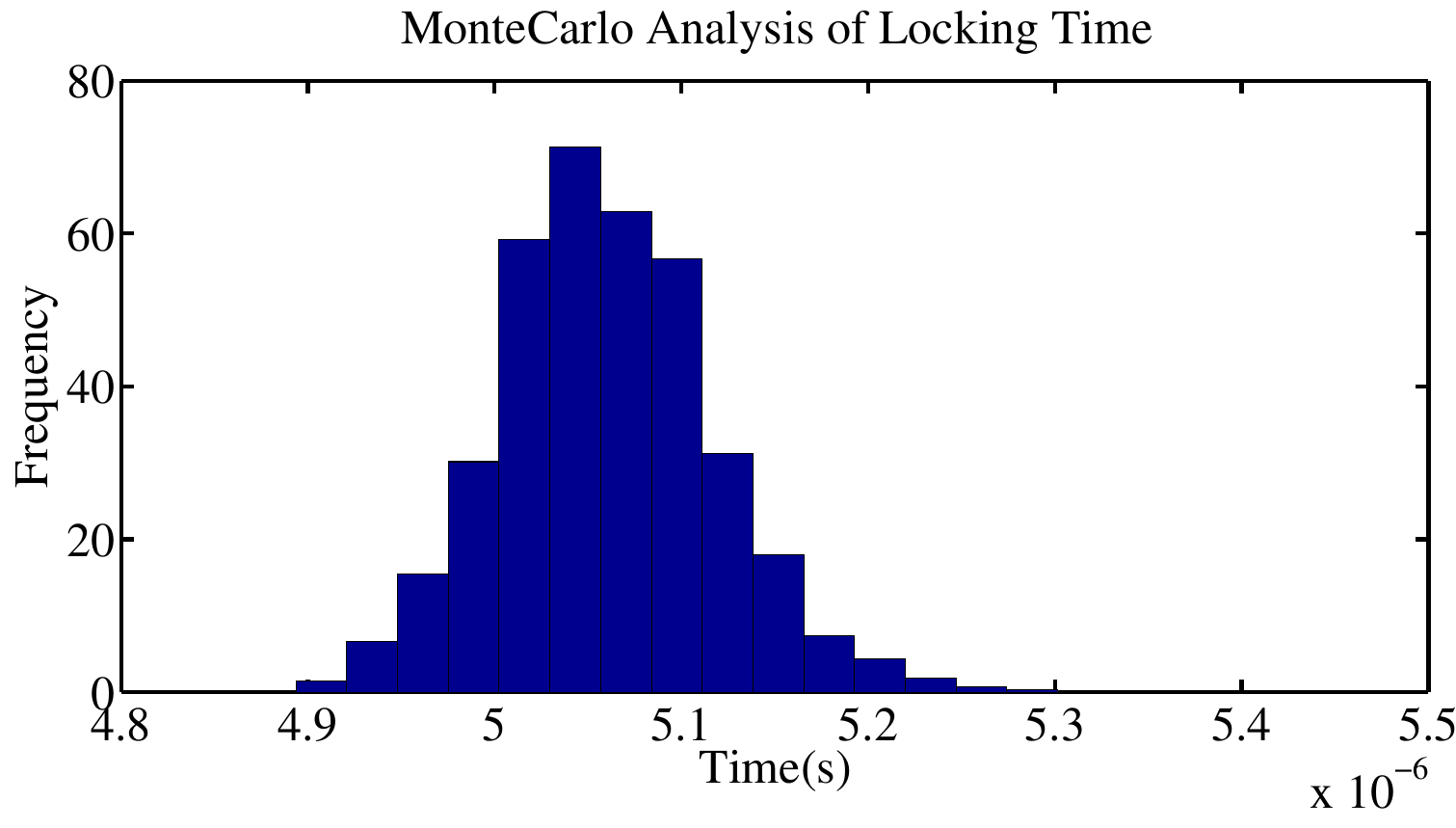}}
\subfigure[Jitter ] {\label{fig:mc_jittKrig}\includegraphics[width=0.45\textwidth]{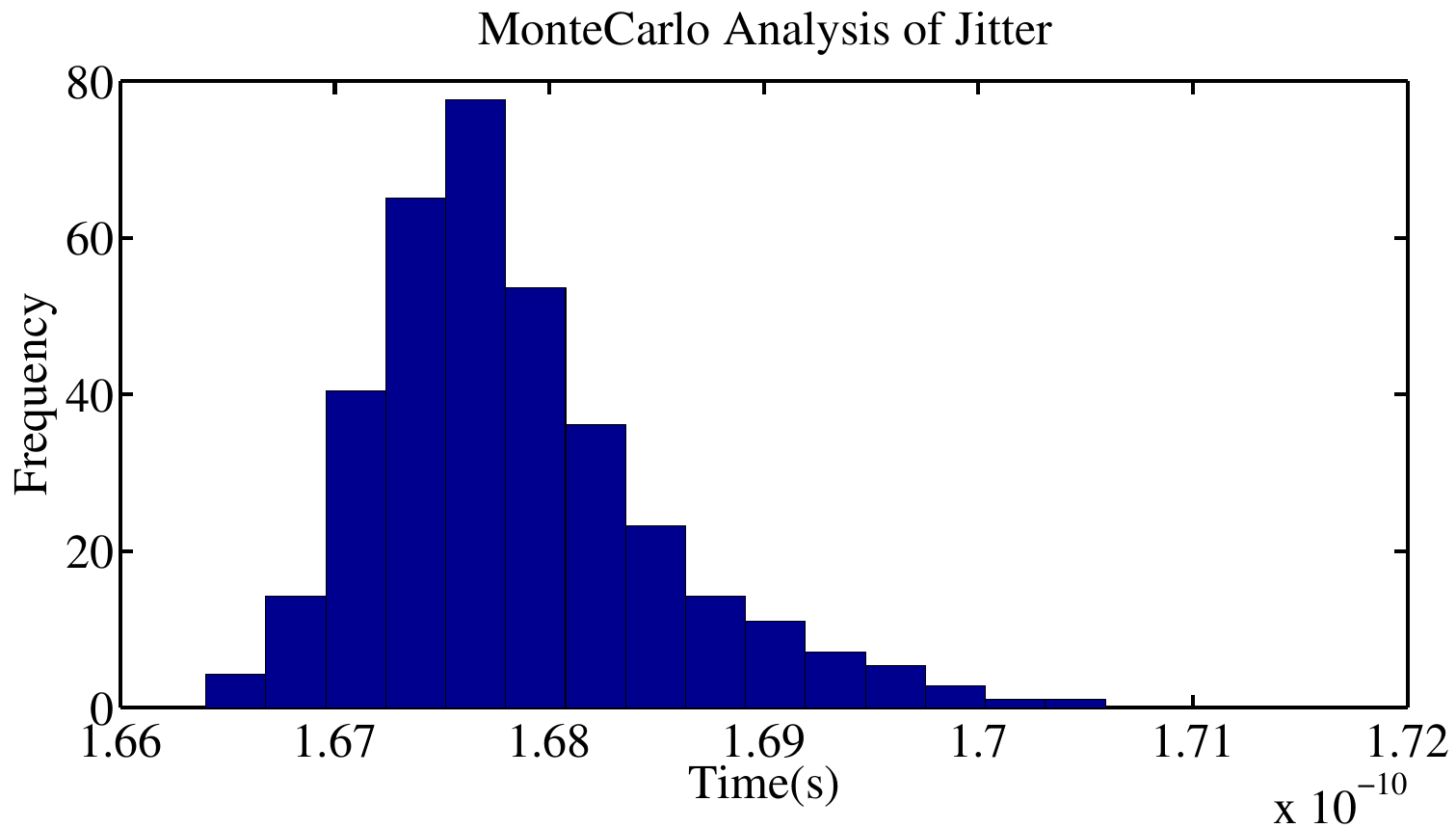}}
\caption{Statistical Analysis of FoMs using Kriging based metamodeling \cite{Okobiah_ISQED_2014}.}
\label{fig:Kriging}
\end{figure*}

\begin{figure*}[hptb]
\centering
\subfigure[Power] {\label{fig:mc_pwrANN}\includegraphics[width=0.45\textwidth]{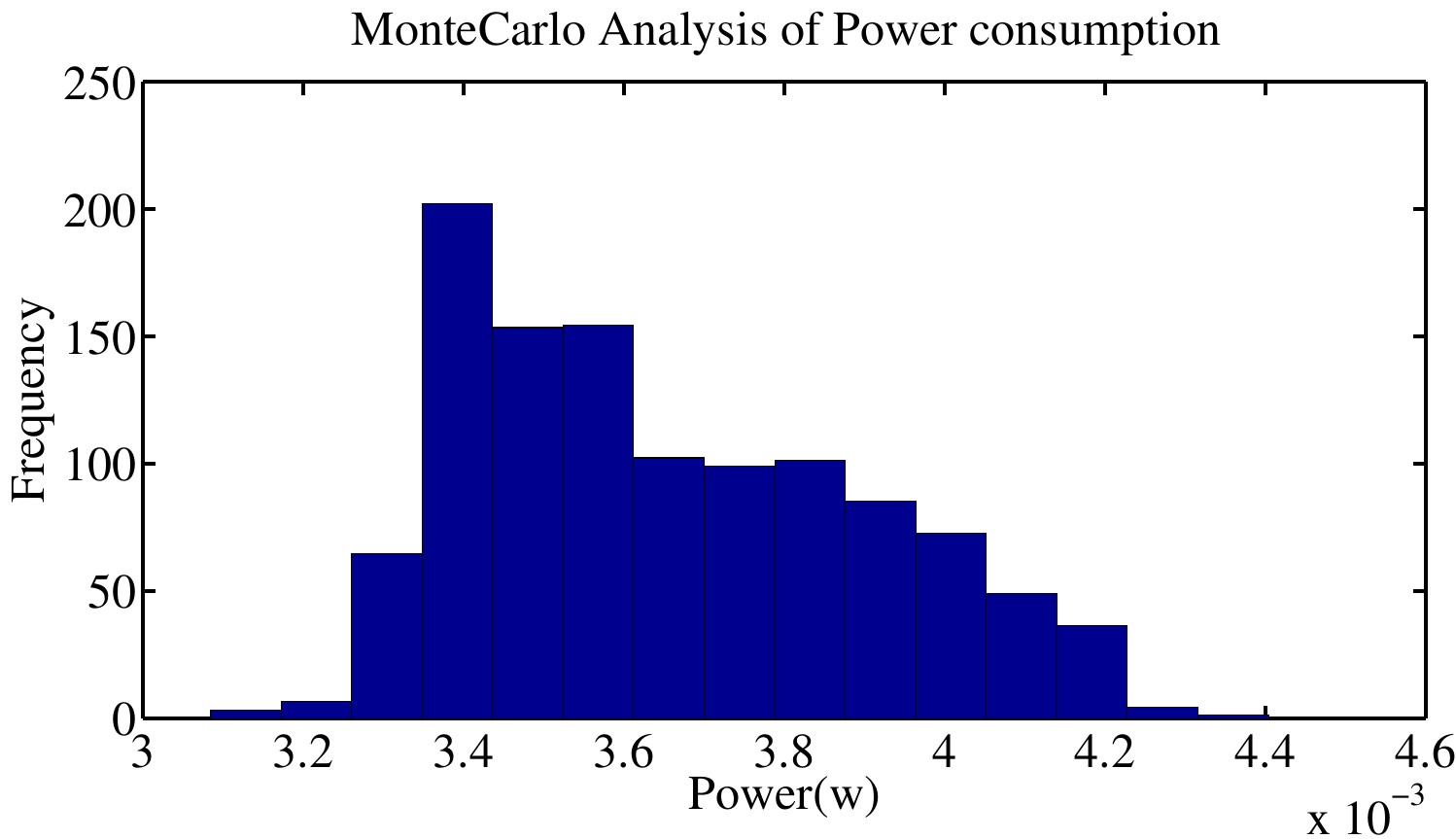}}
\subfigure[Frequency] {\label{fig:mc_freqANN}\includegraphics[width=0.45\textwidth]{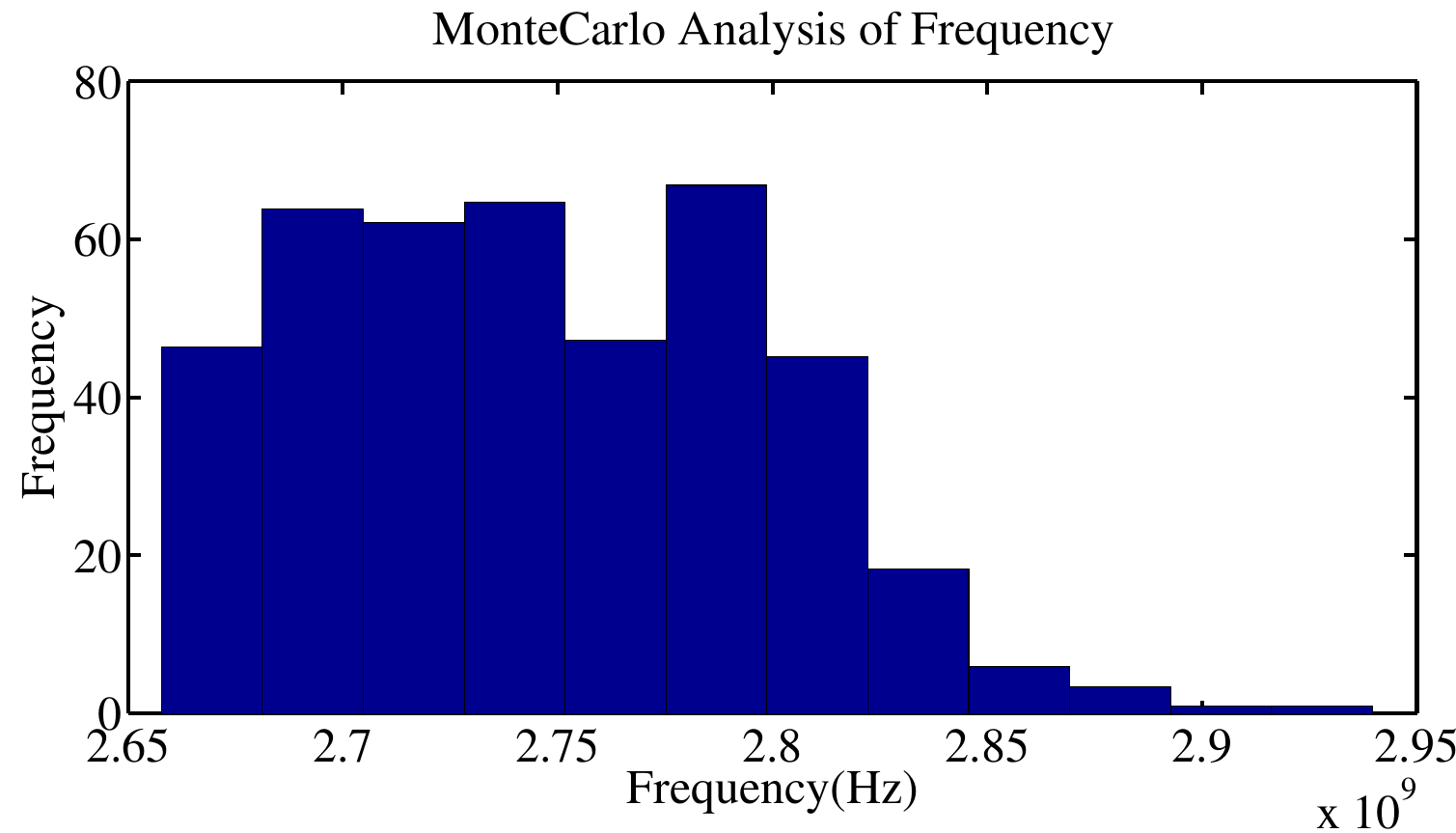}}
\subfigure[Locking Time] {\label{fig:mc_lockANN}\includegraphics[width=0.45\textwidth]{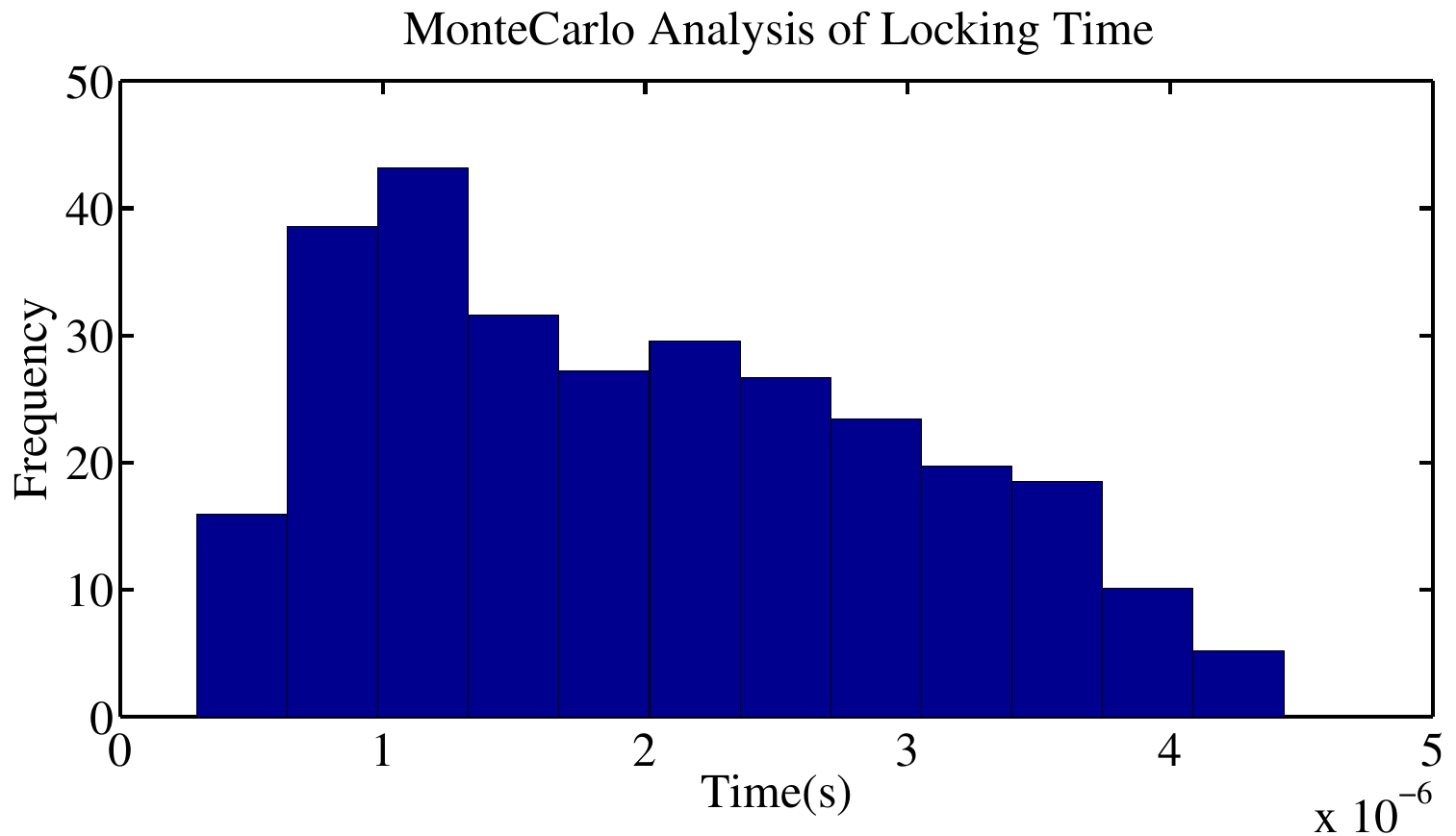}}
\subfigure[Jitter] {\label{fig:mc_jittANN}\includegraphics[width=0.45\textwidth]{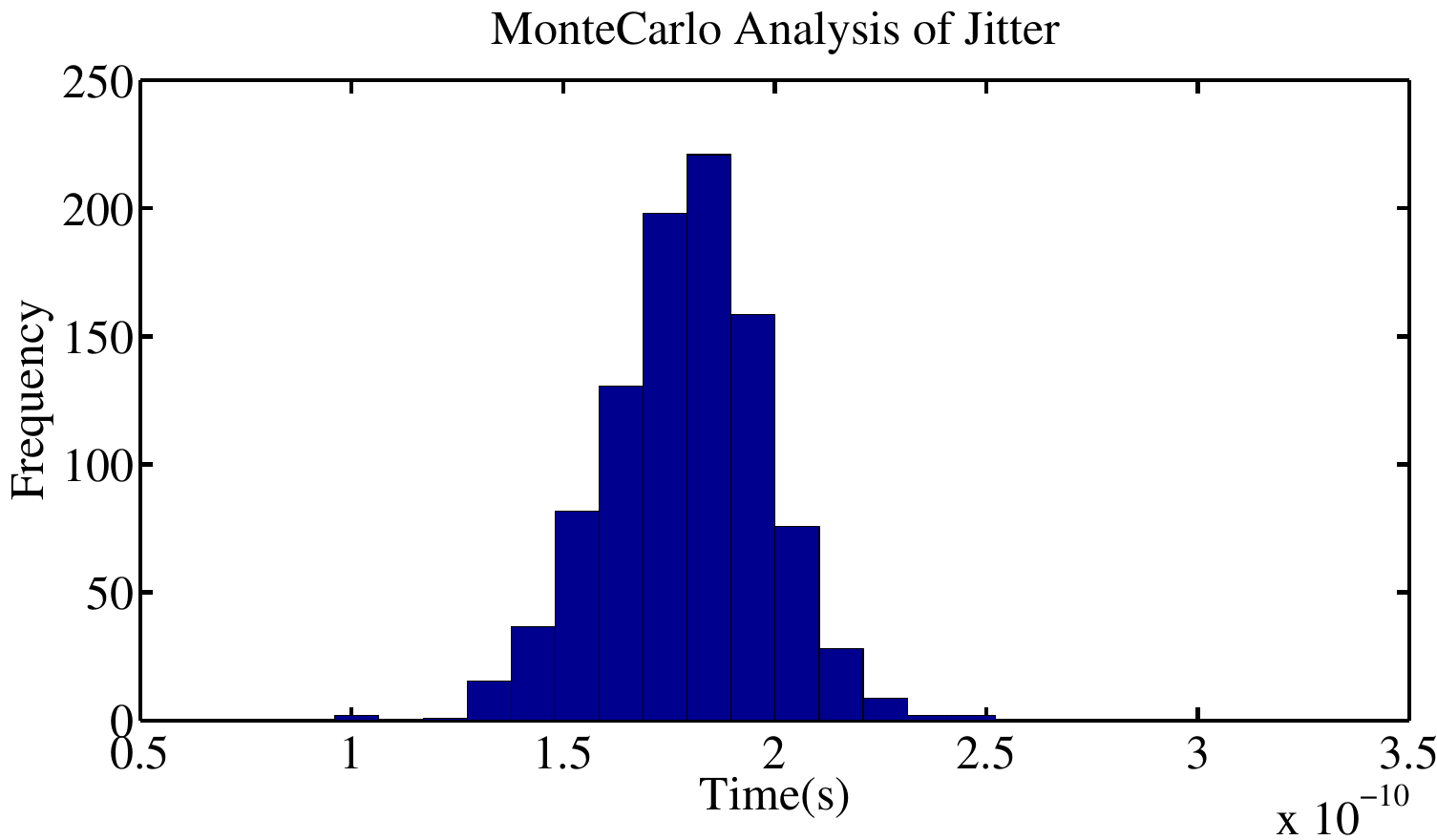}}
\caption{Statistical Analysis of FoMs using Neural Network based metamodeling \cite{Okobiah_ISQED_2014}.}
\label{fig:ANN}
\end{figure*}

Table \ref{Table:ProcessResults} shows the mean $(\mu)$ and standard deviation $(\sigma)$ for the FoMs in each of the metamodels. From the results the Kriging metamodels are shown to be most accurate on both the mean $(\mu)$ and $(\sigma)$ values for all FoMs. The Kriging bootstrapped neural network metamodel on the other hand is shown to be more accurate on the $(\sigma)$ values than the plain neural network metamodel but less accurate on the  $(\mu)$ values. This difference is expected because while bootstrapping infuses the autocorrelation property of Kriging based techniques, some error is also introduced as well. Fig. \ref{fig:mcMean_STD} shows the errors for the $(\mu)$ and $(\sigma)$ as a bar chart. The histograms of the Monte Carlo analysis for the Kriging bootstrapped, Kriging and neural network metamodels are shown in Figures \ref{fig:K_ANN}, \ref{fig:Kriging}, and \ref{fig:ANN}.

The value of the Kriging bootstrapped metamodeling technique is due to the reduced time cost for design exploration. While Kriging models may be more accurate, the time cost for design exploration for a large design space still becomes too expensive due to the repetitive solution of large-dimension systems of equations for \emph{each} sample point. One obvious goal for metamodel use is the improved time cost. Table \ref{Table:TimeAnalysis} shows the time cost for the Monte Carlo Analysis on each metamodel.

\begin{table}[hptb]
\centering
\caption{Monte Carlo Time Analysis Comparison for Metamodels \cite{Okobiah_ISQED_2014}.}
\begin{tabular}{|l|c|c|c|}
	\hline
	Model    & Kriging-ANN  &  Kriging & ANN       \\
	\hhline{|=|=|=|=|} 
	Time     &  19 s        & 468 s    & 19 s      \\  \hline
	Speedup  &  24.63$\times$     & 1        & 24.63$\times$   \\  \hline
\end{tabular}
\label{Table:TimeAnalysis}
\end{table}

 Table \ref{Table:TimeAnalysis} shows a speedup of approximately 25 times in time cost for the Monte Carlo Simulation of 1000 runs for the Kriging bootstrapped model over traditional Kriging. The significant improvement in time cost is large enough to mitigate the minimal error incurred in the metamodel. The overall use of metamodels significantly reduces the simulation time over SPICE models. It may be noted that the Monte Carlo simulation time on the SPICE models is approximately 5 days, which highlights the huge time gain with the use of metamodels.

The experimental analysis was performed using the generated parasitic netlists of the 180nm PLL designs. The optimization of the statistical analysis for the power consumption ($P_{PLL}$) was the objective, while using the locking time ($Lck_{PLL}$) as a design constraint. A total of 21 design parameters were used for the optimization simulation.  Further statistical analysis was carried out using \matlab. The results from the optimization simulation displayed in Table \ref{Table:statResults} show an improved statistical variation of the design simulation. The histograms of the Monte Carlo analysis for the optimized Kriging bootstrapped metamodel are shown in Fig. \ref{fig:K_ANN}.

\begin{table*}[htbp]
\centering
\caption{Statistical Optimization for Kriging Neural Network Metamodel for PLL FoMs \cite{Mohanty_ISQED_2015}.}
\vspace{-5pt}
\begin{tabular}{|l|l|c|c|c|c|c|}
\hline
\multicolumn{2}{|l|}{}    &\multirow{2}{*}{SPICE Netlist}  & \multicolumn{4}{c|}{Kriging-ANN Metamodel} \\ \cline{4-7}
\multicolumn{2}{|l|}{}    &                          & \multicolumn{2}{l|}{Before Optimization}      & \multicolumn{2}{l|}{After Optimization} \\ \cline{3-7}
\multicolumn{2}{|l|}{}    & Value                    & Value       & Error (\%)    & Value       & Error (\%)  \\
\hhline{|==|=|=|=|=|=|} 
\multirow{2}{*}{Power ($P_{PLL}$) }  & Mean ($\mu$)       & 2.48 mW     & 2.40 mW     & 3.33       & 2.35 mW     & 2.08   \\
\cline{2-7}
                             		 & St. Dev. ($\sigma$) & 0.42 mW     & 0.34 mW     & 19.05      & 0.39 mW     & 7.14   \\
\hline
\multirow{2}{*}{Frequency ($F_{PLL}$)} & Mean ($\mu$)       & 2.66 GHz    & 2.51 GHz    & 5.64       & 2.78 GHz    & 4.51    \\
\cline{2-7}
                                       & St. Dev. ($\sigma$) & 10.95 MHz   & 41.93 MHz   & 282.92     & 16.92 MHz    & 54.52   \\
\hline
\multirow{2}{*}{Locking Time ($Lck_{PLL}$)} & Mean ($\mu$)       & 5.51 $\mu$s & 5.11$\mu$s  & 7.26       & 5.21 $\mu$s & 5.44    \\
\cline{2-7}
                                            & St. Dev. ($\sigma$) & 0.72 $\mu$s & 0.44 $\mu$s & 38.88      & 0.42 $\mu$s & 41.67   \\
\hline
\multirow{2}{*}{Jitter ($J_{PLL}$)} & Mean ($\mu$)        & 16.80 ns    & 14.69ns     & 10.25      & 17.72ns     & 5.47   \\
\cline{2-7}
                                    & St. Dev. ($\sigma$)  & 1.32 ps     & 4.50 ps     & 240.91     & 0.33ps      & 75   \\
\hline
\end{tabular}
\label{Table:statResults}
\end{table*}


From the results it is observed that the standard deviation for  $P_{PLL}$, $Lck_{PLL}$, and jitter ($J_{PLL}$) are all minimized with the frequency ($F_{PLL}$) having an increased deviation. The mean power consumption was also reduced while the other FoMs were increased. This is expected since the statistical optimization started off with the design parameters for optimal performance.

\subsection{Comparative Research}
\label{subsec:ComparativeResearch}

Table \ref{Table:ComparsionMetamodels} shows a brief comparison of metamodeling based design techniques. The comparisons are only a perspective and illustrate the applicability and viability of our proposed method for statistical variability analysis. Kriging modeling is presented in \cite{YuICCAD2007}. In \cite{KuoICS2008} a polynomial based metamodeling design including a statistical analysis on process variation is presented.  A polynomial regression based technique is presented in \cite{GaritselovJOLPE12}. The accuracy based on the RMSE value of the models (except for \cite{YuICCAD2007} which uses MSE) is shown in column 4 of Table \ref{Table:ComparsionMetamodels}. 
The presented metamodels have been generated for different circuits, and even when the circuits are similar there are fabricated using different silicon technologies and performance measures making, direct comparisons impossible. Hence, the comparisons are only from a broad perspective point.

\begin{table}[htbp!]
	\centering
	\caption{Comparative Analysis of Related Research \cite{Okobiah_ISQED_2014}.}
	\begin{tabular}{|l|c|c|c|c|}
		\hline
		&                         & Test    &              \\
		Research                              & Metamodel               & Circuit & Accuracy     \\
		\hhline{|=|=|=|=|} 
		Garitselov \cite{GaritselovJOLPE12}   & Polynomial              & PLL     & 0.157       \\  \hline
		\multirow{2}{*}{Yu \cite{YuICCAD2007}}& \multirow{2}{*}{Kriging}& RO      & 0.5325 (MSE)   \\  \cline{3-4}
		&                         & LC-VCO  & 0.5325 (MSE)   \\  \hline
Okobiah \cite{OkobiahISQED2012}       & Kriging                 & Simulated Annealing      & $3.2\times 10^{-9}$   \\  \hline
		Kuo \cite{KuoICS2008}                 & Polynomial              & PLL     & $2.0\times 10^{-4}$    \\  \hline
iVAMS 3.0 \cite{Okobiah_ISQED_2014}             & Kriging-ANN             & PLL     & $2.51\times 10^{-6}$   \\  \hline
	\end{tabular}
\label{Table:ComparsionMetamodels}
\end{table}

\section{Conclusion and Future Research}
\label{sec:Conclusion}


This paper presented a metamodeling design analysis, design exploration and optimization technique that combines traditional Kriging and ANNs to create process aware metamodels. Kriging based techniques are used to bootstrap sample data points which accommodate the correlation characteristics of Kriging techniques into the sample data. Simulation results indeed show an improved process awareness on the metamodels generated for the test case of an 180 nm PLL circuit. The Monte Carlo Simulation time also improved  approximately 25$\times$. The preliminary results are promising. Future research is planned to explore the use of Deep Learning techniques for the ANN structure and training and the automated incorporation of this framework within mixed-signal hardware description languages, such as Verilog-AMS, as presented in our previous work \cite{Mohanty_arXiv_2019-July03-1907-01526_iVAMS2}.

\section*{Acknowledgments}

The current arXiv paper is based on the following conference presentations \cite{Okobiah_ISQED_2014, Mohanty_ISQED_2015}. The current paper presenting iVAMS 3.0 advances the state-of-art of iVAMS 1.0 \cite{Mohanty_arXiv_2019-May31-1905-12812_iVAMS1} and iVAMS 2.0 \cite{Mohanty_arXiv_2019-July03-1907-01526_iVAMS2}, which were presented by us in the past.
\\
\noindent
The authors would like to thank UNT graduate Dr. Oghenekarho Okobiah for his help on conference versions of this work.

\bibliographystyle{IEEEtran}
\bibliography{Bibliography_iVAMS-3-0}

\pagebreak
\section*{Authors' Biographies}
\vspace{-0.2cm}

\begin{wrapfigure}{l}{0.95in}
	\vspace{-0.5cm}
	\includegraphics[width=1.0in,keepaspectratio]{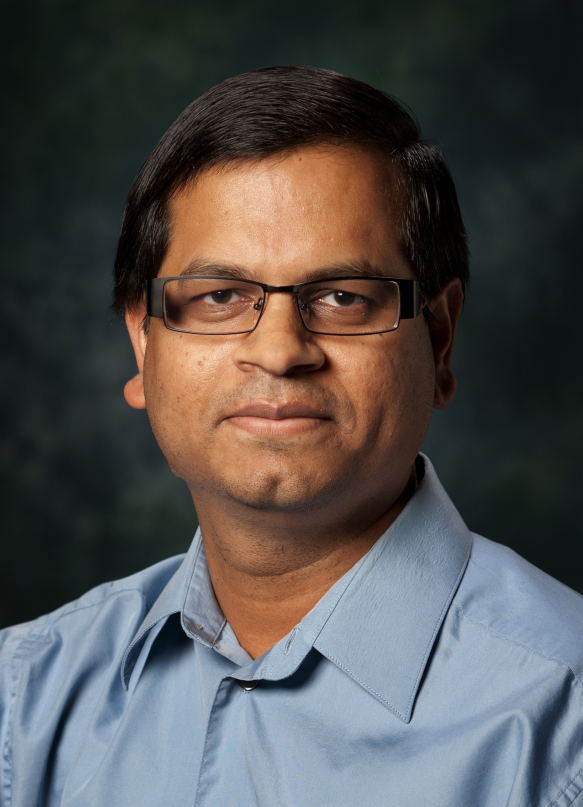}
	\vspace{-0.9cm}
\end{wrapfigure}
\textbf{Saraju P. Mohanty} received the bachelor’s degree (Honors) in electrical engineering from the Orissa University of Agriculture and Technology, Bhubaneswar, in 1995, the master’s degree in Systems Science and Automation from the Indian Institute of Science, Bengaluru, in 1999, and the Ph.D. degree in Computer Science and Engineering from the University of South Florida, Tampa, in 2003. He is a Professor with the University of North Texas. His research is in ``Smart Electronic Systems’’ which has been funded by National Science Foundations (NSF), Semiconductor Research Corporation (SRC), U.S. Air Force, NIDILRR, IUSSTF, and Mission Innovation. He has authored 550 research articles, 5 books, and 10 granted and pending patents. His Google Scholar h-index is 62 and i10-index is 300 with 17,000 citations. He is regarded as a visionary researcher on Smart Cities technology in which his research deals with security and energy aware, and AI/ML-integrated smart components. He introduced the Secure Digital Camera (SDC) in 2004 with built-in security features designed using Hardware Assisted Security (HAS) or Security by Design (SbD) principle. He is widely credited as the designer for the first digital watermarking chip in 2004 and first the low-power digital watermarking chip in 2006. He is a recipient of 21 best paper awards, Fulbright Specialist Award in 2021, IEEE Consumer Electronics Society Outstanding Service Award in 2020, the IEEE-CS-TCVLSI Distinguished Leadership Award in 2018, and the PROSE Award for Best Textbook in Physical Sciences and Mathematics category in 2016. He has delivered 31 keynotes and served on 15 panels at various International Conferences. He has been serving on the editorial board of several peer-reviewed international transactions/journals, including IEEE Transactions on Big Data (TBD), IEEE Transactions on Computer-Aided Design of Integrated Circuits and Systems (TCAD), IEEE Transactions on Consumer Electronics (TCE), and ACM Journal on Emerging Technologies in Computing Systems (JETC). He has been the Editor-in-Chief (EiC) of the IEEE Consumer Electronics Magazine (MCE) during 2016-2021. He served as the Chair of Technical Committee on Very Large Scale Integration (TCVLSI), IEEE Computer Society (IEEE-CS) during 2014-2018 and on the Board of Governors of the IEEE Consumer Electronics Society during 2019-2021. He serves on the steering, organizing, and program committees of several international conferences. He is the steering committee chair/vice-chair for the IEEE International Symposium on Smart Electronic Systems (IEEE-iSES), the IEEE-CS Symposium on VLSI (ISVLSI), and the OITS International Conference on Information Technology (OCIT). He has supervised 3 post-doctoral researchers, 19 Ph.D. dissertations, 29 M.S. theses, and 41 undergraduate projects.

\vspace{0.4cm}
\begin{wrapfigure}{l}{0.95in}
	\vspace{-0.5cm}
	\includegraphics[width=1.0in,keepaspectratio]{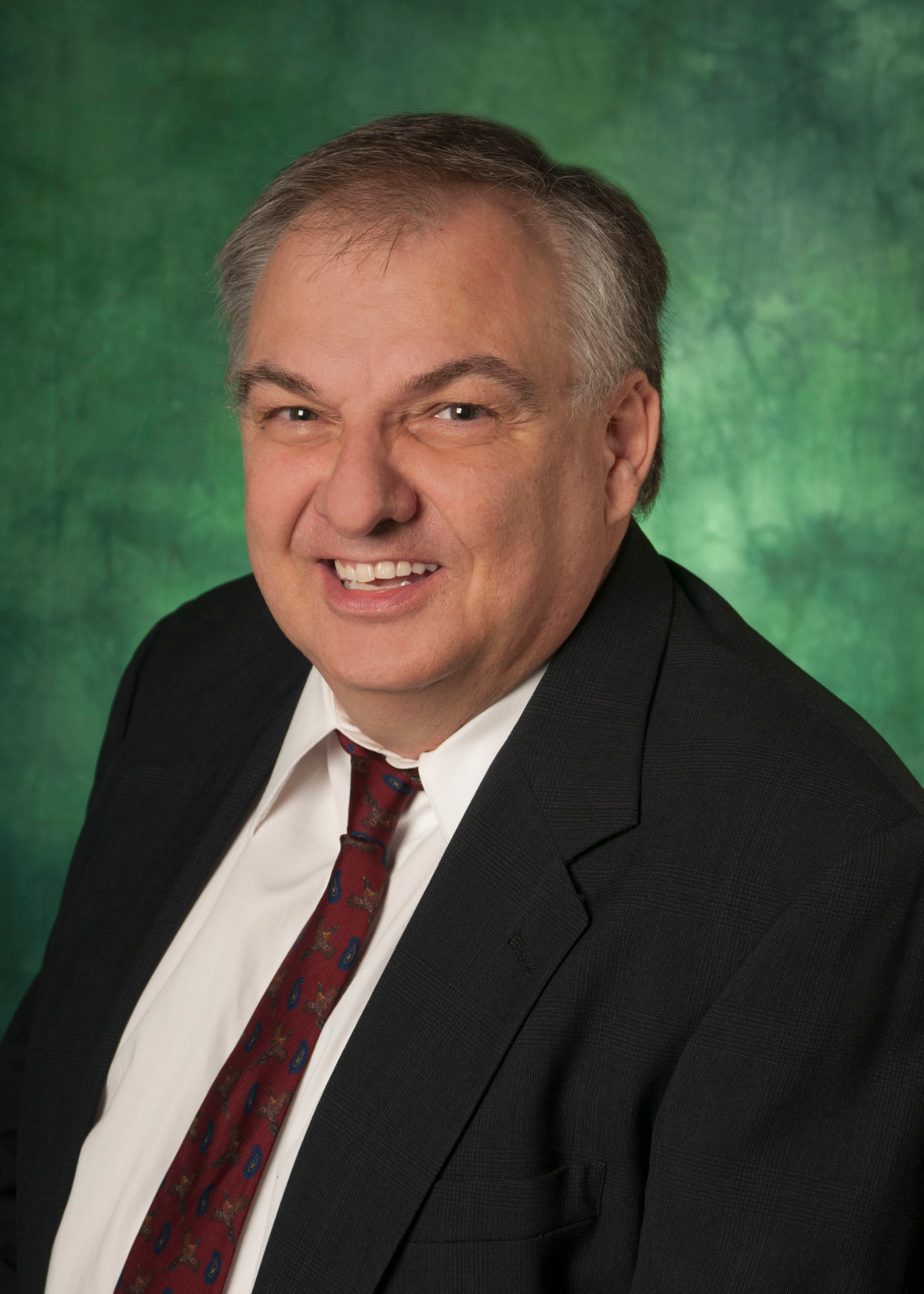}
	\vspace{-0.9cm}
\end{wrapfigure}
\noindent
\textbf{Elias Kougianos}  received a BSEE from the University of Patras, Greece in 1985 and an MSEE in 1987, an MS in Physics in 1988 and a Ph.D. in EE in 1997, all from Louisiana State University. From 1988 through 1998 he was with Texas Instruments, Inc., in Houston and Dallas, TX. In 1998 he joined Avant! Corp. (now Synopsys) in Phoenix, AZ as a Senior Applications engineer and in 2000 he joined Cadence Design Systems, Inc., in Dallas, TX as a Senior Architect in Analog/Mixed-Signal Custom IC design. He has been at UNT since 2004. He is a Professor in the Department of Electrical Engineering, at the University of North Texas (UNT), Denton, TX. His research interests are in the area of Analog/Mixed-Signal/RF IC design and simulation and in the development of VLSI architectures for multimedia applications. He is an author of over 200 peer-reviewed journal and conference publications.

\end{document}